\documentclass[final,3p,times]{elsarticle}

\usepackage{amssymb}
\usepackage{amsmath}
\usepackage{caption}
\usepackage{graphicx}
\usepackage{float} 
\usepackage{subfigure}
\usepackage{subcaption}
\usepackage{booktabs} 
\usepackage{color}
\usepackage{multirow}
\usepackage{xurl}
\usepackage[hidelinks]{hyperref}

\journal{Neurocomputing}

\begin{document}

\begin{frontmatter}



    \title{MCI-GRU: Stock Prediction Model Based on Multi-Head Cross-Attention and Improved GRU}

    \author[label1]{Peng Zhu}
    \author[label1]{Yuante Li}
    \author[label1]{Yifan Hu}
    \author[label2]{Sheng Xiang}
    \author[label1]{Qinyuan Liu \corref{cor1}}
    \author[label1]{Dawei Cheng}
    \author[label3]{Yuqi Liang}
    \cortext[cor1]{Corresponding author}
    \affiliation[label1]{organization={Department of Computer Science and Technology, Tongji University},
        city={Shanghai},
        country={China}}
    \affiliation[label2]{organization={Australian Artificial Intelligence Institute, University of Technology Sydney},
        city={Sydney},
        country={Australia}}
    \affiliation[label3]{organization={Seek Data Group, Emoney Inc.},
        city={Shanghai},
        country={China}}

    \begin{abstract}
        As financial markets become increasingly complex and the era of big data unfolds, accurate stock prediction has become more critical. Although traditional time series models, such as GRU, have been widely applied to stock prediction, they still exhibit limitations in addressing the intricate nonlinear dynamics of markets, particularly in the flexible selection and effective utilization of key historical information. In recent years, emerging methods like Graph Neural Networks and Reinforcement Learning have shown significant potential in stock prediction. However, these methods often demand high data quality and quantity, and they tend to exhibit instability when dealing with data sparsity and noise. Moreover, the training and inference processes for these models are typically complex and computationally expensive, limiting their broad deployment in practical applications. Existing approaches also generally struggle to capture unobservable latent market states effectively, such as market sentiment and expectations, microstructural factors, and participant behavior patterns, leading to an inadequate understanding of market dynamics and subsequently impact prediction accuracy. To address these challenges, this paper proposes a stock prediction model, MCI-GRU, based on a multi-head cross-attention mechanism and an improved GRU. First, we enhance the GRU model by replacing the reset gate with an attention mechanism, thereby increasing the model's flexibility in selecting and utilizing historical information. Second, we design a multi-head cross-attention mechanism for learning unobservable latent market state representations, which are further enriched through interactions with both temporal features and cross-sectional features. Finally, extensive experiments conducted on the CSI 300 and CSI 500 datasets from the Chinese stock market, as well as the NASDAQ 100 and S\&P 500 datasets from the U.S. stock market, demonstrate that the proposed method outperforms the current state-of-the-art methods across multiple metrics. Furthermore, this approach has been successfully applied in the real-world operations of a fund management company, validating its effectiveness and practicality in actual financial environments. The code is available at \url{https://github.com/WinstonLiyt/MCI-GRU}.
    \end{abstract}


\end{frontmatter}


\section{Introduction}
In recent years, with the advent of the big data era and the rapid development of the global economy, the complexity of financial markets \cite{greenwood1997financial} has significantly increased. This trend has posed unprecedented challenges to the volatility and unpredictability of stock markets \cite{bustos2020stock}. Consequently, accurate stock prediction \cite{singh2017stock} has become critically important not only for investors and financial institutions, enabling them to formulate more robust investment strategies and risk management measures, but also for policymakers, who rely on these predictions for macroeconomic regulation and market oversight. Additionally, for academic researchers, stock prediction has emerged as a pivotal domain for uncovering market dynamics and behavioral patterns, thereby advancing the study of financial market theories and data-driven methodologies. These investigations not only extend the theoretical boundaries of financial economics but also provide new research directions and application scenarios for interdisciplinary fields such as machine learning and data science \cite{raschka2020machine}. Therefore, the accuracy and efficacy of stock prediction have become focal points across multiple disciplines, further stimulating extensive exploration into innovative models and methods.

Time series models \cite{liu2021forecast, cheng2018modeling}, such as GRU and LSTM \cite{gao2021stock}, have been widely utilized in stock prediction due to their significant advantages in capturing temporal dependencies within sequential data. However, these models exhibit limitations when addressing long-term dependencies in financial markets. Long-term trends and large-scale fluctuations are often obscured by noise, making it challenging for these models to extract valuable long-term dependency information from such noisy data effectively. Additionally, financial markets are characterized by a high degree of nonlinearity, with rapid shifts in market behavior driven by changes in investor sentiment, unexpected events, and other factors. These models often lack sufficient sensitivity and flexibility in handling such nonlinearity and abrupt events. Furthermore, they face challenges in the flexible selection and effective utilization of critical historical information. Given the vast and disorganized nature of data in financial markets, identifying the most relevant features for prediction has become a critical issue.

In recent years, the Transformer model \cite{han2021transformer} has demonstrated significant potential in capturing long-range dependencies and handling complex nonlinear features, owing to its architecture based on self-attention mechanisms \cite{ding2022novel}. Unlike traditional RNN models such as GRU and LSTM, the Transformer can simultaneously attend to all time steps within a sequence, making it particularly effective in extracting dependencies over extended time spans. Furthermore, the Transformer's strong parallel processing capabilities enable it to efficiently manage large-scale data, which is especially crucial when dealing with vast and diverse stock data in financial markets. However, the application of Transformer models also presents several challenges. First, the large number of parameters in Transformer models leads to high computational costs, particularly when processing ultra-large-scale financial data, potentially limiting their application in resource-constrained environments. Additionally, while the Transformer is adept at capturing complex nonlinear relationships, its performance may be compromised when confronted with highly noisy financial data. Therefore, to fully harness the advantages of the Transformer, it is often necessary to incorporate more sophisticated preprocessing and feature selection methods to enhance its accuracy and efficiency in stock prediction tasks.

Artificial intelligence technologies' rapid advancement \cite{goralski2020artificial}, particularly in Graph Neural Networks (GNNs) \cite{wu2020comprehensive, han2025mitigating} and Reinforcement Learning (RL) \cite{moerland2023model}, has introduced unprecedented potential for stock prediction. These technologies, through innovative algorithmic designs and deep learning models, promise to enhance the ability to capture the complex dynamics of financial markets. For example, methods that employ GNNs are able to precisely capture the complex and diverse interdependencies within financial data by modeling the relationships between stocks as graph structures. This approach not only uncovers deep connections that are difficult for traditional models to reveal but also more effectively reflects the nonlinear characteristics of the market. Meanwhile, methods based on RL progressively learn and optimize trading strategies by simulating continuous interaction with the market environment. This method is highly adaptive, enabling dynamic strategy adjustments to cope with the market's rapidly changing conditions. However, despite the considerable promise of these emerging methodologies, they still face significant challenges in practical application. Firstly, these models often rely on large-scale, high-quality datasets, which are difficult to acquire or construct in real-world scenarios. When confronted with data sparsity and noise—common issues in financial markets—the predictive performance of these models can be severely compromised. Additionally, the computational complexity of these methods is high; the training process is not only time-consuming and resource-intensive but also demands substantial computational power, significantly limiting their widespread adoption in practical financial applications. More critically, a fundamental limitation of current models lies in their inability to capture unobservable latent market states effectively. Factors such as market sentiment, investor expectations, microstructural elements, and participant behavior patterns play crucial roles in shaping market dynamics. However, the failure to adequately account for these latent factors often leads to a superficial understanding of the market, thereby constraining the predictive accuracy and practical utility of these models.

To address these challenges, this paper introduces a novel stock prediction model, MCI-GRU, which integrates a multi-head cross-attention mechanism and improved GRU architecture. First, by replacing the reset gate in the traditional GRU model with an attention mechanism, the MCI-GRU model significantly improves the flexibility in selecting and utilizing historical time series information. Second, MCI-GRU employs a Graph Attention Network (GAT) to extract cross-sectional features from stock data. Additionally, this paper introduces a multi-head cross-attention mechanism designed to capture latent, unobservable market states. The model's expressive capability and ability to capture complex market dynamics are further enhanced by interacting these latent states with temporal and cross-sectional features. We conduct extensive experimental evaluations on several stock market datasets, including the CSI 300 and CSI 500 indices in China, as well as the NASDAQ 100 and S\&P 500 indices in the United States. The experimental results demonstrate that the proposed MCI-GRU model outperforms existing state-of-the-art methods across multiple key performance metrics. Furthermore, the model has been successfully deployed in the operations of a leading fund management company, showcasing its practical applicability and effectiveness in real-world financial environments. In summary, the key contributions of this paper are as follows:
\begin{itemize}

    \item This paper improves the traditional GRU model by replacing the reset gate with an attention mechanism, enabling the model to more flexibly select critical historical information, thereby enhancing the effectiveness of filtering and utilizing past sequence data.

    \item We propose a multi-head cross-attention mechanism to learn representations of unobservable market latent states. These learned representations are then interactively integrated with both temporal and cross-sectional features, thereby enriching the model's feature representation capacity.

    \item We conduct empirical studies on stock market datasets from multiple countries, and the results demonstrate that the proposed method outperforms existing state-of-the-art approaches. Moreover, the method has been successfully deployed in practical applications on a fund company's platform.
\end{itemize}

\section{Related Work}
Stock market prediction has been a long-standing challenge in the field of finance, commonly solutions include traditional and machine learning, deep and reinforcement learning, graph neural networks, and the latest methods.

\subsection{Traditional Learning and Machine Learning Methods}
Traditional approaches, including Autoregressive~(AR) \cite{autoregressive}, ARIMA \cite{ARIMA, 7046047}, and Exponential Smoothing \cite{exponsmoothing, de2009predicting}, have been extensively employed in stock prediction, primarily for modeling linear trends. With advancements in computational technology, machine learning methods such as Hidden Markov Models(HMM) \cite{HMM, HMM-base}, Support Vector Machines~(SVM) \cite{SVM, SVM2}, K-Nearest Neighbor~(KNN) \cite{guo2003knn, alkhatib2013stock}, Decision Trees \cite{DecisionTree}, and Neural Networks \cite{NN, NN-2, NN-3}, including Single Layer Perceptron~(SLP) and MultiLayer Perceptron~(MLP) \cite{pan2005predicting, atsalakis2009surveying}, have garnered significant attention for their ability to model complex patterns in stock data. For instance, \cite{8250694} introduced a decision tree-based approach demonstrating the efficacy of Random Forests in short-term prediction and the superior long-term accuracy of J48 combined with Bagging. Similarly, \cite{ZHANG20191} utilized high-order Hidden Markov Models with advanced parameters, including state transition probabilities dependent on multiple previous states and observation probabilities modeled as Gaussian mixtures. This method incorporated a state dimension reduction technique to simplify parameter estimation and decoding, alongside a dynamic trading strategy based on identified hidden states, validated on the CSI 300 and S\&P 500 indices. Moreover, \cite{chen2017feature} integrated feature-weighted support vector machine~(FWSVM) and feature-weighted k-nearest neighbor~(FWKNN) techniques by calculating feature importance via information gain, which informed weight assignment in SVM classification and distance calculation in KNN.

While these advanced machine learning techniques have demonstrated their capacity to capture complex nonlinear interactions within stock data, challenges remain, including susceptibility to overfitting due to the low signal-to-noise ratio, high trading volumes, frequent trading, significant price volatility, and the multitude of influencing factors inherent in financial markets.

\subsection{Deep Learning and Reinforcement Learning Methods}
With the rapid advancement of deep learning, this technology has been extensively applied to stock prediction in financial markets, producing notable outcomes \cite{dl1, NN-2}. Recurrent Neural Networks (RNNs) \cite{akita2016deep, karim2022stock, LSTM-base} have demonstrated exceptional capabilities in this field by effectively modeling long-term dependencies in time series data, utilizing inputs such as stock prices to forecast market trends. Recent studies have introduced sophisticated models aimed at further enhancing prediction accuracy. For instance, \cite{gupta2022stocknet} proposed the StockNet model based on GRU, which incorporates an injection module to mitigate overfitting and a survey module for comprehensive stock analysis. Furthermore, \cite{CLVSA} integrated convolutional LSTM units with a sequence-to-sequence framework and attention mechanisms, employing variational methods and backward decoders to improve prediction accuracy and robustness. Similarly, \cite{ALSTM-base} advanced the attention-based LSTM model by implementing adversarial training to enhance its generalization capability.

Deep learning models often demonstrate instability when confronted with extreme market fluctuations, such as those experienced during the 2008 financial crisis \cite{farmer2012stock} and the 2019 COVID-19 pandemic \cite{he2020impact}. In response, reinforcement learning models \cite{han2023efficient} have gained prominence due to their adaptability and capacity for continuous learning. Reinforcement learning approaches in investment strategies can be broadly categorized into two types: value-based and policy-based \cite{sun2023reinforcement, yang2020deep}. Value-based approaches involve learning a critique to estimate the expected outcomes of trading actions within the market. Common value-based approaches in investment strategies include Q-learning \cite{neuneier1995optimal}, Deep Q-learning \cite{carta2021multi, jin2016portfolio}, Recurrent Reinforcement Learning \cite{deng2016deep, ding2018investor}, and Sarsa \cite{corazza2015q, de2020tabular}. However, a major limitation of value-based approaches is the complexity of accurately approximating the market environment with a critic. As a result, policy-based approaches \cite{daberius2019deep, jiang2017cryptocurrency} are often regarded as more suitable for financial markets. For example, \cite{AlphaStock} integrated deep attention networks with reinforcement learning, optimizing parameters through discrete agent actions to maximize the Sharpe ratio of investments. To further balance profit and risk, \cite{si2017multi} introduced a multi-objective deep reinforcement learning (MODRL) approach for intraday trading of stock index futures, combining deep learning for feature extraction with reinforcement learning for decision-making. Despite their potential, reinforcement learning models face challenges, such as the requirement for large datasets and difficulties in model interpretability, which can hinder their practical application in financial markets.

\subsection{Graph Neural Networks and Latest Methods}
In recent years, Graph Neural Networks \cite{cheng2025graph, wang2022review, cheng2022financial} have garnered significant attention in stock prediction due to their capacity to capture complex interdependencies within financial data. For instance, \cite{REST} proposed a hybrid model that integrates Recurrent Neural Networks with GNN, facilitating real-time predictions. Similarly, \cite{THGNN} introduced a hierarchical attention mechanism into GNN, thereby improving the model's ability to analyze multi-level market dependencies and perform structured analyses of stock trends. To capture a stock's intrinsic value more accurately, \cite{qiao2023higherorder} developed a Higher-order Graph Attention Network (H-GAT), which differentiates itself by modeling complex subgraph structures involving more than two stocks and incorporating both technical and fundamental factors. This approach contrasts with traditional GNNs, which typically consider only simple pairwise relationships, thus enhancing the model's ability to reflect the intrinsic value of stocks. However, many graph-based models frequently overlook the diversity of stock price changes and the temporal dynamics inherent in these fluctuations, necessitating the development of more innovative graph-based approaches. For instance, \cite{MASTER} introduced a market-guided stock transformer that can dynamically simulate the instantaneous and cross-temporal correlations among stocks, leading to enhanced accuracy in stock trend predictions. Additionally, \cite{CSM} successfully integrated long-term trends, short-term fluctuations, and sudden events into a cohesive graph-based framework, significantly surpassing traditional methods by accounting for the multi-scale nature of market dynamics.
\cite{liu2024echo} proposed ECHO-GL, a model that leverages earnings call-derived relations and multimodal graph learning to predict stock movements and generate temporal price trajectories.
\cite{qian2024mdgnn} proposed the Multi-relational Dynamic Graph Neural Network (MDGNN), which integrates discrete dynamic graph modeling with Transformer-based temporal encoding to capture evolving stock relationships, demonstrating superior performance over state-of-the-art methods in stock movement prediction.
\cite{zhao2022stock} proposed the DANSMP model for stock movement prediction, which utilizes a Market Knowledge Graph comprising bi-typed entities and hybrid relations, combined with Dual Attention Networks to effectively capture momentum spillover signals and enhance predictive accuracy.
Despite their promising capabilities in stock prediction, GNNs exhibit several notable limitations. These models often struggle to effectively capture complex nonlinear relationships and account for anomalous market scenarios, which are crucial in financial forecasting. Moreover, GNN-based models tend to be sensitive to data sparsity and noise, which can undermine their robustness, especially in real-world financial datasets where missing or unreliable data is common. Additionally, certain graph-based approaches, such as those proposed by \cite{liu2024echo}, \cite{qian2024mdgnn}, and \cite{zhao2022stock}, may face challenges in generalization due to their reliance on specific data structures and sophisticated attention mechanisms, which can lead to overfitting, particularly in environments with limited or imbalanced training data. These models, while effective in their respective domains, often struggle with scalability and flexibility when applied to diverse market conditions and broader datasets, limiting their applicability in more volatile or unpredictable markets. Thus, enhancing the robustness and generalizability of graph-based models remains a key area for improvement in stock movement prediction.

With the rapid advancements in Large Language Models (LLMs) \cite{achiam2023gpt, brown2020language, touvron2023llama}, their application in stock prediction has garnered significant scholarly interest. LLMs, such as GPT-4, have demonstrated remarkable capabilities in natural language understanding, making them highly suitable for financial sentiment analysis and predictive modeling. Research by \cite{lopezlira2023} revealed a strong correlation between sentiment analysis generated by ChatGPT for news headlines and subsequent daily stock market returns, highlighting the potential of LLMs in capturing market sentiment and its impact on stock prices. This study underscores the utility of LLMs in extracting and quantifying sentiment from unstructured textual data, which can be critical for short-term stock forecasting. Moreover, the integration of LLMs with Graph Neural Networks has opened new avenues for enhancing stock prediction accuracy. For example, \cite{Chen_2023} employed ChatGPT to infer dynamic network structures from financial news, which were subsequently incorporated into a GNN for stock prediction. This approach not only leverages the linguistic prowess of LLMs in understanding and summarizing complex financial information but also capitalizes on the GNN’s ability to model intricate relationships between stocks. The resulting hybrid model demonstrated superior predictive performance, suggesting that the synergistic combination of LLMs and GNNs can effectively address the challenges of dynamic and interconnected financial markets.

\begin{figure}[t]
    \centering
    \includegraphics[width=1.0\textwidth]{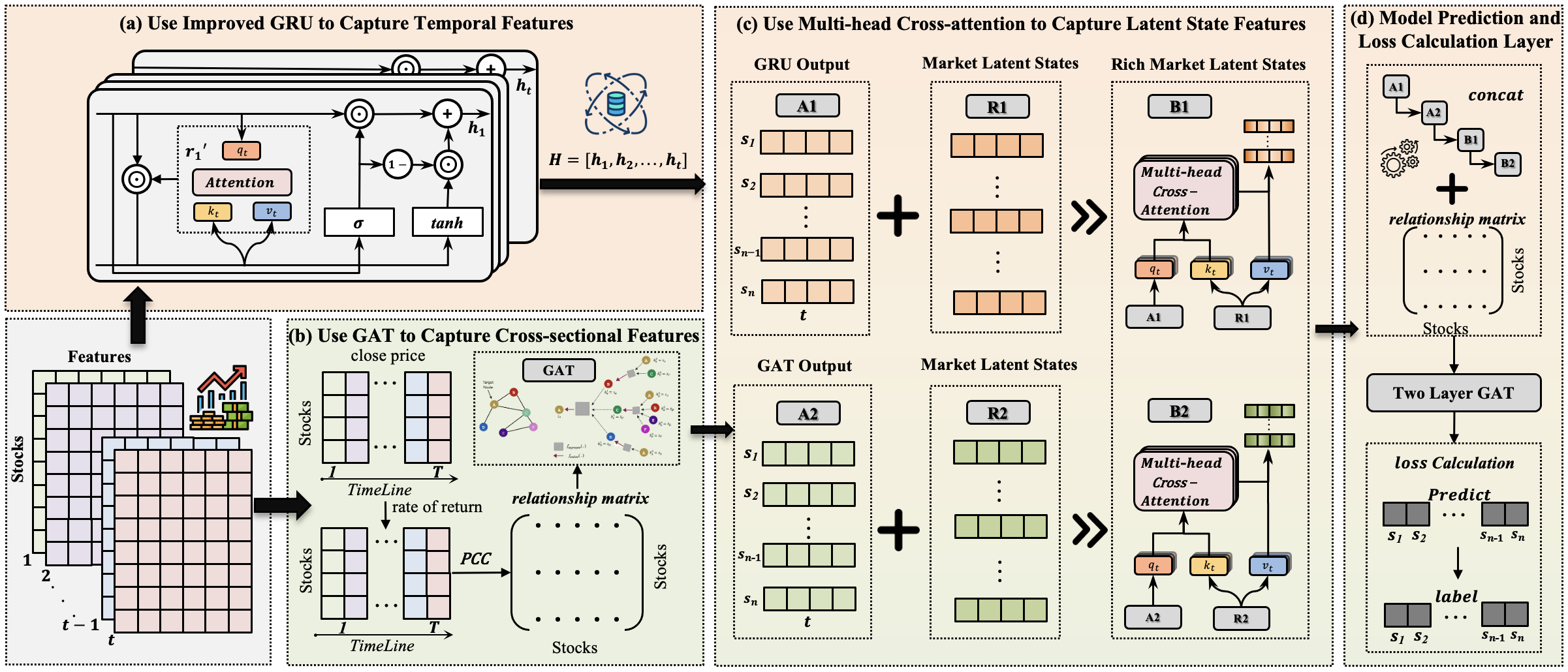}
    \caption{The architecture of the proposed model MCI-IGRU.
        The enhanced GRU model in Part (a) integrates an attention mechanism in place of the reset gate, greatly improving its ability to capture and learn from temporal patterns. Part (b) leverages attention mechanisms to identify and weigh relationships between stocks, effectively extracting cross-sectional features. Part (c) captures latent market conditions influencing stock behavior, allowing the model to learn and represent hidden, non-observable market states. In part (d), the final prediction process integrates these learned features, refining outcomes through optimized loss calculation for improved performance.}
    \label{figure1}
\end{figure}

\section{Methodology}
In this section, we give a detailed introduction to the MCI-GRU model proposed in this paper and the model structure is shown in Figure \ref{figure1}. The whole model is divided into four modules: Use Improved GRU to Capture Temporal Features~(Part a), Use GAT to Capture Cross-sectional Features~(Part b), Use Multi-head Cross-attention to Capture Latent State Features~(Part c), Model Prediction and Loss Calculation Layer~(Part d). Part(a) employs an enhanced GRU model that replaces the reset gate with an attention mechanism, significantly enhancing the model's capacity to represent and learn from temporal data. Part(b) enhances the model by using attention mechanisms to capture and weigh the relationships between different stocks, thereby extracting cross-sectional features from the data. Part(c) captures hidden market conditions that affect stock behavior, enabling the model to learn and characterize latent, non-directly observable market state features. Part(d) refines the final predictions by integrating the learned features and calculating the loss to optimize model performance. In the following subsections, we first give the pre-definition and then introduce each module in detail.

\subsection{Predefinition}
We consider a collection of stocks denoted by the set $S= \left \{s_{1},s_{2},\cdots,s_{N} \right \}$, where each $s_{i}$ represents an individual stock, and $N$ is the total number of stocks within the dataset. For any stock $s_{i}$, the data associated with the $t$-th trading day is represented by the vector $x_{it} = \left \{x_{it}^{open},x_{it}^{close},x_{it}^{high},x_{it}^{low},x_{it}^{volume},x_{it}^{turnover} \right \}$, where $x_{it}^{open}$, $x_{it}^{close}$, $x_{it}^{high}$, $x_{it}^{low}$, $x_{it}^{volume}$ and $x_{it}^{turnover}$ correspond to the opening price, closing price, highest price, lowest price, trading volume, and turnover amount on day $t$, respectively. We let $d_{x}$ denote the number of features used to describe each stock on each day. The time series data for stock $s_{i}$ over $t$ days is encapsulated in the set $x_{i} = \left \{x_{i1},x_{i2},\dots,x_{it} \right \}$. Collectively, the dataset for all stocks is represented as $X = \left \{x_{1},x_{2},\dots,x_{N} \right \}$.

\subsection{Use Improved GRU to Capture Temporal Features}
In time series prediction tasks, the GRU model has been widely employed due to its capability to capture temporal dependencies within sequential data effectively. However, traditional GRU models have certain limitations in capturing complex temporal relationships, particularly when dealing with long-term dependencies, where they may struggle to extract deeper features from the sequence. To address this, the present work utilizes an enhanced GRU model, incorporating an attention mechanism in place of the reset gate, thereby improving the model's ability to represent and learn from temporal data.

\subsubsection{Basic Structure of the GRU}
In the classical GRU model, the hidden state $h_{t}$ is updated through two gating mechanisms: the update gate $z_{t}$ and the reset gate $r_{t}$. The formulas are as follows:
\begin{equation}
    \centering
    \begin{split}
        z_{t} = \sigma(W_{z}x_{t} + U_{z}h_{t-1}+b_{z})\\
        r_{t} = \sigma(W_{r}x_{t} + U_{r}h_{t-1}+b_{r})
    \end{split}
    \centering
\end{equation}
where $x_{t} \in \mathbb{R}^{d_{x}}$ denotes the input of the stock at the current time step, $h_{t-1} \in \mathbb{R}^{d_{h}}$ represents the hidden state from the previous time step and $d_{h}$ represents the dimension of the hidden state. $W_{z} \in \mathbb{R}^{d_{h} \times d_{x}}$, $U_{z} \in \mathbb{R}^{d_{h} \times d_{h}}$, $W_{r} \in \mathbb{R}^{d_{h} \times d_{x}}$ and  $U_{r} \in \mathbb{R}^{d_{h} \times d_{h}}$ are weight matrices. $b_{z} \in \mathbb{R}^{d_{h}}$ and $b_{r} \in \mathbb{R}^{d_{h}}$ are bias terms, and $\sigma$ is the activation function, typically the sigmoid function. $z_{t} \in \mathbb{R}^{d_{h}}$ and the reset gate $r_{t} \in \mathbb{R}^{d_{h}}$ is employed to regulate the influence of the previous time step’s hidden state $h_{t-1}$ when computing the candidate's hidden state $\tilde{h_{t}} \in \mathbb{R}^{d_{h}}$ at the current time step, as detailed by the following formula:
\begin{equation}
    \centering
    \begin{split}
        \tilde{h_{t}} = \tanh(W_{h}x_{t} + r_{t}\odot(U_{h}h_{t-1})+b_{h})
    \end{split}
    \centering
\end{equation}
where $W_{h} \in \mathbb{R}^{d_{h} \times d_{x}}$, $U_{h} \in \mathbb{R}^{d_{h} \times d_{h}}$ are weight matrices. $b_{h} \in \mathbb{R}^{d_{h}}$ is bias terms and $\odot$ denotes element-wise multiplication. Ultimately, the hidden state $h_{t}$ at the current time step is controlled by the update gate $z_{t}$, as shown below:
\begin{equation}
    \centering
    \begin{split}
        h_{t} = (1-z_{t})\odot h_{t-1} + z_{t}\odot \tilde{h_{t}}
    \end{split}
    \centering
\end{equation}
In these equations, the reset gate $r_{t}$ determines how much the previous time step’s hidden state $h_{t-1}$ is ``reset'' at the current time step. However, this mechanism exhibits certain limitations in capturing long-term dependencies.

\subsubsection{Introduction of the Improved GRU}
To overcome the aforementioned limitations, the proposed model introduces an attention mechanism to replace the traditional reset gate $r_{t}$ in GRU. The attention mechanism dynamically allocates weights to different time steps in the sequence, thereby capturing critical information within the time series data more precisely. Specifically, the traditional reset gate $r_{t}$ is replaced with an attention-based weight coefficient $\alpha_{t}$, computed as follows:
\begin{equation}
    \centering
    \begin{split}
        \alpha_{t} = Attention(h_{t-1}, x_{t})
    \end{split}
    \centering
\end{equation}
where $\alpha_{t} \in \mathbb{R}$ is the attention weight vector. The core idea of the attention mechanism is to allocate attention weights by calculating the similarity between the query, key, and value. In this model, the hidden state $h_{t-1} \in \mathbb{R}^{d_{h}}$ from the previous time step is treated as the query, while the current time step's input $x_{t}$ serves as both the key and value.
Attention calculation includes three parts: Linear Transformations of Query, Key, and Value, Weighted Sum, and Attention Weight Calculation. The specific implementation and mathematical formula are as follows:
\begin{itemize}
    \item Linear Transformations of Query, Key, and Value: First, the hidden state $h_{t-1}$ and the input $x_{t}$ are linearly transformed into query, key, and value spaces respectively:
          \begin{equation}
              \centering
              \begin{split}
                  q_{t} = W_{q}h_{t-1}, \ \ \ \
                  k_{t} = W_{k}x_{t}, \ \ \ \
                  v_{t} = W_{v}x_{t}
              \end{split}
              \centering
          \end{equation}
          where $W_{q} \in \mathbb{R}^{d_{q} \times d_{h}}$, $W_{k} \in \mathbb{R}^{d_{k} \times d_{h}}$ and $W_{v} \in \mathbb{R}^{d_{v} \times d_{h}}$  are learnable linear transformation matrices. $q_{t} \in \mathbb{R}^{d_{q}}$ and $d_{q}$ represent the dimensions of the query, which is used to calculate the attention weight. $k_{t} \in \mathbb{R}^{d_{k}}$ and $d_{k}$ represents the dimensions of key.  $v_{t} \in \mathbb{R}^{d_{v}}$ and $d_{v}$ represents the dimension of value, which is usually the same as $d_q$ and $d_{k}$, that is, $d_{q}=d_{k}=d_{v}$.

    \item Attention Weight Calculation: The attention weights are obtained by computing the dot product similarity between the query and key:
          \begin{equation}
              \centering
              \begin{split}
                  \alpha_{t} = softmax(\frac{q_{t}{k}^{\top}_{t}}{\sqrt{d_{k}}})
              \end{split}
              \centering
          \end{equation}
          where $d_{k}$ is used to scale the dot product results to prevent numerical instability.

    \item Weighted Sum: The final reset gate value is derived from a weighted sum of the values, using the attention weights:
          \begin{equation}
              \centering
              \begin{split}
                  r_{t}^{'} = \alpha_{t}v_{t}
              \end{split}
              \centering
          \end{equation}
          where the new $r_{t}^{'} \in \mathbb{R}^{d_{v}}$ dynamically selects the most important parts of the current input $x_{t}$ and the previous hidden state $h_{t-1}$, thereby enhancing the model's ability to capture long-term dependencies.
\end{itemize}

\subsubsection{Update of the Hidden State}

The calculation of the update gate $z_{t}$ remains unchanged. With the new reset gate $r_{t}^{'}$, the update formula for the GRU’s hidden state is adjusted as follows:
\begin{equation}
    \centering
    \begin{split}
        \tilde{h_{t}^{'}} = \tanh(W_{h}x_{t} + r_{t}^{'}\odot(U_{t}h_{t-1})+b_{h}) \\
        h_{t} = (1-z_{t})\odot h_{t-1} + z_{t}\odot \tilde{h_{t}^{'}}
    \end{split}
    \centering
\end{equation}
where $z_{t}$ is the update gate. This updated hidden state calculation integrates the dynamic information selection capability provided by the attention mechanism, allowing the model to better extract long-term dependency information and key features from the time series data.

\subsubsection{Final Output}
Through recursive computation over multiple time steps, the enhanced GRU model generates a final sequence of hidden states $H=[h_{1},h_{2},\dots,h_{t}]$, where each $h_{t}$ incorporates information from past time steps, with increased focus on important time steps due to the attention mechanism. For subsequent processing, the final hidden state $h_{t}$ is typically taken as the representation vector for the entire sequence, denoted as $A_{1} \in \mathbb{R}^{N \times d_{h}}$. This output $A_{1}$ is then used as input for further feature extraction and model learning in the next processing stage.

\subsection{Use GAT to Capture Cross-sectional Features}
The Graph Attention Network is a key component in the model architecture, responsible for extracting cross-sectional features from the data by capturing the relationships between different stocks. GAT extends the traditional Graph Convolutional Network by incorporating attention mechanisms, allowing the model to assign different levels of importance to different nodes (stocks) in the graph based on their relationships.

\subsubsection{Input Representation}
In this model, the input to the GAT layer is a matrix representing the features of all stocks at a specific time step. The input matrix has dimensions $(N, d_{x})$, $N$ is the number of stocks that represent the nodes in the graph, and $d_{x}$ is the dimensionality of the feature vector for each stock. This input representation is derived from the origin stock data. Hence, the GAT layer particularly focuses on capturing the cross-sectional dependencies between stocks.

\subsubsection{Graph Construction}
The graph in the GAT layer is constructed where each node represents a stock, and the edges between nodes represent the relationships between these stocks. The edges' weights, or the relationships' strengths, are determined by the historical correlations of the stocks’ returns over the past year. These correlations are typically computed using Pearson correlation or other statistical measures. We calculate the historical correlation of stock returns over the past year for any two stocks $s_{i}$ and $s_{j}$ to determine the strength of their relationship. First, let $r_{i}(t')$ and $r_{j}(t')$ represent the returns of stocks $s_{i}$ and $s_{j}$ at time $t$, where $t' = 1, 2, \dots, T$ denotes the number of trading days in the past year (e.g., 252 trading days per year). The returns are typically calculated using the log return formula:
\begin{equation}
    \centering
    \begin{split}
        r_{i}(t') = \frac{x_{it'}^{close}-x_{i(t'-1)}^{close}}{x_{i(t'-1)}^{close}}
    \end{split}
    \centering
\end{equation}
where $r_{i}(t')$ represents the closing price of the stock $s_{i}$ at time $t'$. Next, we compute the Pearson correlation coefficient $\rho(s_{i}, s_{j})$ between the return series of the two stocks, which measures their linear correlation. The Pearson correlation coefficient is given by the following formula:
\begin{equation}
    \centering
    \begin{split}
        \rho(s_{i}, s_{j}) = \frac{{\textstyle \sum_{t'=1}^{T}(r_{i}(t')-\bar{r_{i}})(r_{j}(t')-\bar{r_{j}} )} }{\sqrt{{\textstyle \sum_{t'=1}^{T}(r_{i}(t')-\bar{r_{i}})^{2}}}\sqrt{{\textstyle \sum_{t'=1}^{T}(r_{j}(t')-\bar{r_{j}})^{2}}} }
    \end{split}
    \centering
\end{equation}
where $\bar{r_{i}}$ and $\bar{r_{j}}$ represent the average returns of stocks $s_{i}$ and $s_{j}$ over the past year.
Based on this correlation coefficient $\rho(s_{i}, s_{j})$, we assign an edge weight to represent the strength of the relationship between stocks $s_{i}$ and $s_{j}$. The weight $w_{i,j}$ is typically set to $\rho(s_{i}, s_{j})$. To optimize the learning process, not all relationships are included in the graph. Instead, threshold-based filtering is applied using a parameter known as $judge_{value}$. This parameter allows the model to retain only edges that represent significant relationships, effectively reducing noise and focusing the model on the most relevant connections.

\subsubsection{Attention Mechanism in GAT}
The core of the GAT layer lies in its attention mechanism, which dynamically computes the importance (attention coefficients) of each node’s neighbors when aggregating information. For each node $i$ in the graph, the GAT layer performs the following operations:
\begin{itemize}
    \item Linear Transformation: Each node’s feature vector $x_{it}$ is linearly transformed using a learnable weight matrix $W_{g}$:
          \begin{equation}
              \centering
              \begin{split}
                  h_{i}^{'}  = W_{g}x_{it}
              \end{split}
              \centering
          \end{equation}
          where $x_{it}$ is the date of the stock $s_{i}$ in time $t$, $W_{g} \in \mathbb{R}^{d_{g}\times d_{x}}, $ $h_{i}^{'} \in \mathbb{R}^{d_{g}}$ is the transformed feature vector. $d_{g}$ is the dimension of the hidden layer.

    \item Attention Coefficients Calculation: The attention coefficients between stock
          $i$ and its neighbor $j$ are computed using the following equation:
          \begin{equation}
              \centering
              \begin{split}
                  e_{ij}^{} = LeakyReLU(a^{\top}[h_{i}^{'}||h_{j}^{'}])
              \end{split}
              \centering
          \end{equation}
          where $a \in \mathbb{R}^{2d_{g}}$ is a learnable attention vector, $||$ denotes concatenation, $LeakyReLU$
          is a non-linear activation function that introduces non-linearity to the attention computation.

    \item Normalization: The attention coefficients are then normalized across all neighbors of node $i$ using the softmax function:
          \begin{equation}
              \centering
              \begin{split}
                  \sigma_{ij} = \frac{\exp(e_{ij})}{\sum_{k \in N(i)}^{}exp(e_{ik})}
              \end{split}
              \centering
          \end{equation}
          where $\sigma_{ij}$ is the normalized attention score between stock $i$ and stock $j$, $N_{i}$ denotes the set of neighbors of the node $i$.

    \item Feature Aggregation: Finally, the node’s updated representation is computed as a weighted sum of its neighbors’ transformed features, weighted by the normalized attention coefficients:
          \begin{equation}
              \centering
              \begin{split}
                  h_{i}^{''} = \sigma(\sum_{j \in N(i)}^{}\sigma_{ij}h_{j}^{'})
              \end{split}
              \centering
          \end{equation}
          where $\sigma$ is a non-linear activation function, typically ReLU.
\end{itemize}

\subsubsection{Output of the GAT Layer}
The final output of the GAT layer is a matrix $A_{2} \in \mathbb{R}^{N \times d_{g}}$, where each row corresponds to the updated feature vector of stock, now enriched with information from its neighbors. This output captures the cross-sectional dependencies between stocks and serves as the input to the next stage of the model, where the market latent state learning layer further processes these features.

\subsection{Use Multi-head Cross-attention to Capture Latent State Features}
The Market Latent State Learning Layer is a crucial component of the model, designed to capture and represent the latent states of the market that are not directly observable from the raw data. This layer is specifically tailored to model the underlying market conditions that influence stock behavior, allowing the model to better understand and predict stock movements by leveraging these hidden states.

\subsubsection{Initialization of Market Latent States}
The process begins with the initialization of a set of learnable market latent state vectors. These vectors are meant to represent different latent market conditions or factors that could be driving stock prices. The initialization is as follows:
\begin{itemize}
    \item Number of Latent States ($d_r$): The model initializes $d_r$ latent state vectors, where $d_r$ is a hyperparameter. This parameter determines the number of latent states, and it is typically adjusted based on the complexity of the modeled market. A larger $d_r$ allows the model to capture a wider variety of potential market factors, thereby enhancing its representational capacity. However, a $d_r$ that is too large could increase computational costs and introduce the risk of overfitting. Therefore, selecting an appropriate $d_r$ value is crucial, and it is usually obtained through cross-validation or experimental adjustments to achieve the best performance. In this paper, we set it to $16$.

    \item Latent State Dimension ($d_{i}$): The dimension of each latent state vector is $d_{i}$, which should match the dimensionality of the stock features learned by the model in previous layers such as the GRU and GAT layers. Ensuring that $d_{i}$ is consistent with the output dimensions of the earlier layers allows these latent state vectors to interact effectively with other layers. The choice of $d_{i}$ typically depends on the needs of the model and the characteristics of the data. A larger $d_{i}$ provides more representational power but may increase computational complexity. We set it to $32$ In this paper.
\end{itemize}

The initialized latent state vectors are denoted as $R_1 \in \mathbb{R}^{d_{h} \times d_{i}} $ and $R_2 \in \mathbb{R}^{d_{h} \times d_{i}}$, corresponding to the two main types of information processed by the model: time series information (from the improved GRU output $A_1$ and cross-sectional information (from the GAT output $A_2$. These two types of information play different roles in the model, ensuring that the model can fully capture the dynamic changes in the market. The dimensions of $R_1$ and $R_2$ are both $(d_r, d_i)$, ensuring their representational capacity and enabling effective interaction with subsequent network layers.

\subsubsection{Multi-head Cross-Attention Mechanism}
The core of the market latent state learning process involves a multi-head cross-attention mechanism. This mechanism allows the latent state vectors to interact with the outputs from the Improved GRU and GAT layers, effectively absorbing the relevant information from these outputs and refining the latent state representations.

\textbf{Multi-head Cross-Attention between $R_{1}$ and $A_{1}$:}
In the multi-head cross-attention mechanism, the interaction between $A_{1}$ and $R_{1}$ can be modeled through the cross-attention process, where $A_{1}$ is used as the query and $R_{1}$ is used as the key and value.
Using $A_{1}$ as the query and $R_{1}$ as both the key and value follows the natural logic of information processing, where $A_{1}$ guides the update of $R_{1}$, allowing effective integration of time series information into the latent state. Using $R_{1}$ as the query is logically flawed for two reasons: the information flow is reversed, as $A_{1}$ should influence $R_{1}$, and $R_{1}$, being randomly initialized, lacks a clear objective as a query target. The goal is to refine $R_{1}$ using $A_{1}$, not the other way around.
Here's how the multi-head cross-attention mechanism operates between $A_{1}$ and $R_{1}$:
\begin{itemize}
    \item Linear Transformations: For each attention head $i$, we compute the attention scores between $A_{1}$(Query) and $R_{1}$(Key and Value) as follows:
          \begin{equation}
              \centering
              \begin{split}
                  Q_{i} = A_{1}W_{i}^{Q}, \ \
                  K_{i} = R_{1}W_{i}^{K}, \ \
                  V_{i} = R_{1}W_{i}^{V}
              \end{split}
              \centering
          \end{equation}
          where $A_{1} \in \mathbb{R}^{N \times d_{h}}$, $R_{1} \in \mathbb{R}^{d_{r} \times d_{i}}$ and we set $d_{h}=d_{i}$. For head $i$, $W_{i}^{Q} \in \mathbb{R}^{d_{h} \times d_{h'}}$ is the learnable weight matrix for queries, $W_{i}^{K} \in \mathbb{R}^{d_{h} \times d_{h'}}$ is the learnable weight matrix for keys, $W_{i}^{V} \in \mathbb{R}^{d_{h} \times d_{h'}}$ is the learnable weight matrix for values. $Q_{i} \in \mathbb{R}^{N \times d_{h'}}$, $K_{i} \in \mathbb{R}^{d_{r} \times d_{h'}}$ and $V_{i} \in \mathbb{R}^{d_{r}  \times d_{h'}}$. $d_{h'} = \frac{d_{h}}{h'}$ is the dimension of each attention head and $h'$ is the number of attention heads.

    \item Scaled Dot-Product Attention: For each head $i$, we compute attention scores between the query $A_{1}$ and the key $R_{1}$ as follows:
          \begin{equation}
              \centering
              \begin{split}
                  head_{i} = Attention(Q_{i},K_{i},V_{i}) = softmax(\frac{Q_{i}K_{i}^{\top}}{\sqrt{d_{h'}}})V_{i}
              \end{split}
              \centering
          \end{equation}
          where $Q_{i}K_{i}^{\top} \in \mathbb{R}^{N \times d_{r} }$ gives the attention weights, $\frac{1}{\sqrt{d_{h'}}}$ is the scaling factor to avoid overly large dot-product values. The softmax function ensures that the attention weights sum to 1.

    \item Concatenation of Attention Heads: After computing attention for each of the $h'$ heads, the outputs of all heads are concatenated:
          \begin{equation}
              \centering
              \begin{split}
                  B_{1} = MultiHead(A_{1},R_{1}) = Concat(head_{1},\dots,head_{h'})W^{O}
              \end{split}
              \centering
          \end{equation}
          where $W^{O}$ is a learnable matrix used to project the concatenated results into the output space. The final output after applying the multi-head cross-attention mechanism between $A_{1}$ and $R_{1}$ is $B_{1} \in \mathbb{R}^{N \times d_{h}}$, which captures complex relationships between the two sets of features.

    \item  Calculation and update process of $R_{1}$: During the model training process, the update of $R_1$ is achieved by calculating the gradient of the loss function with respect to $R_1$ through backpropagation. First, the model generates predictions through a forward pass and calculates the loss between the predictions and the actual values. Common loss functions include Mean Squared Error. Through the backpropagation algorithm, we can compute the gradient of the loss function $L$ with respect to $R_1$, i.e., $ \frac{\partial L}{\partial R_1} $. Since $R_1$ affects the output of subsequent layers, the gradient of the loss function is propagated back to $R_1$ through the chain rule. During backpropagation, the gradients of the loss function with respect to $K_i$ and $ V_i$ will propagate back to $R_1$, and the gradient computation is as follows:
          \begin{equation}
              \centering
              \begin{split}
                  \frac{\partial L}{\partial R_1} = \sum_{i=1}^{h'} \frac{\partial L}{\partial K_i} \cdot \frac{\partial K_i}{\partial R_1} \\
                  \frac{\partial L}{\partial R_1} = \sum_{i=1}^{h'} \frac{\partial L}{\partial V_i} \cdot \frac{\partial V_i}{\partial R_1}
              \end{split}
              \centering
          \end{equation}
          In practice, the gradient is propagated through the keys ($K_i$) and values ($V_i$) of each attention head and finally aggregated back to $R_1$. The gradient is computed based on the linear dependencies $K_i = R_1 W_i^K$ and $ V_i = R_1 W_i^V$. Once the gradient of the loss function with respect to $R_1 $ is calculated, the model updates $R_1 $ using the gradient descent algorithm. The update rule is:
          \begin{equation}
              \centering
              \begin{split}
                  R_1 = R_1 - \eta \cdot \frac{\partial L}{\partial R_1}
              \end{split}
              \centering
          \end{equation}
          where $ \eta $ is the learning rate, controlling the step size of the parameter update. Through this process, $ R_1 $ is adjusted according to the gradient calculated by backpropagation, allowing it to more effectively represent the latent market states and capture the complex patterns in the data.
\end{itemize}

\textbf{Multi-head Cross-Attention between $R_{2}$ and $A_{2}$:} The second multi-head cross-attention operation is performed similarly, but this time between the latent state vectors $R_{2}$ and the GAT output $A_{2}$. By a similar method, we calculated $B_{2} \in \mathbb{R}^{N \times d_{g}}$.

\subsubsection{Integration of Market Latent States}
After the multi-head cross-attention mechanisms have been applied, the outputs $B_{1}$ and $B_{2}$ are considered to be enriched latent state representations. These vectors now capture the essential time series and cross-sectional features of the market, making them powerful representations for subsequent tasks, such as predicting stock movements or identifying market regimes.

\subsection{Model Prediction and Loss Calculation Layer}
The Loss Calculation Layer is the final stage of the model and is responsible for synthesizing the outputs from previous layers and generating predictions. This layer also defines how the model is trained by computing the difference between the predicted values and the ground truth, which is then minimized during the training process.

\subsubsection{Input Composition}
The inputs to the Loss Calculation Layer are the outputs from three key components of the model: the time-series representation $A_{1}$, the cross-sectional representation $A_{2}$, and the latent market state representations $B_{1}$ and $B_{2}$. Time-Series Representation $A_{1}$ comes from the improved GRU layer, which captures time-series dependencies in the stock data. Cross-sectional representation $A_{2}$ comes from the GAT layer, which models the relationships between stocks based on their inter-stock correlations. Latent Market State Representations $B_{1}$ and $B_{2}$ come from the Market Latent State Learning Layer, representing the hidden market factors learned from the time series and cross-sectional data.

To prepare for the final prediction, the model concatenates these outputs to form a comprehensive feature vector that integrates all relevant information. The concatenation can be expressed as:
\begin{equation}
    \centering
    \begin{split}
        Z=Concat(A_{1}, A_{2}, B_{1}, B_{2})
    \end{split}
    \centering
\end{equation}
where $Z \in \mathbb{R}^{N \times d_{z}}$ is the combined feature vector that will be used for the final prediction, and $d_{z} = 2d_{h}+2d_{g}$ depends on the individual dimensions of $A_{1}$, $A_{2}$, $B_{1}$ and $B_{2}$.

\subsubsection{Final Prediction with GAT Layers}
Once the feature vector $Z$ is obtained, it is passed through additional GAT layers to make the final prediction. The purpose of these layers is to refine the concatenated features by considering the relationships between stocks (nodes) in a graph structure, as GAT layers are well-suited to modeling graph-structured data. The GAT layers in this part of the model function similarly to the earlier GAT layers, but they now operate on a more comprehensive set of features, combining both temporal and cross-sectional information, as well as the latent market state representations. The structure of the GAT layers can be described as follows:
\begin{itemize}
    \item Graph Construction: The graph’s structure is the same as in the earlier GAT layer. The nodes represent individual stocks, and the edges represent relationships between the stocks based on their historical correlations over the past year. These correlations are filtered using the judge\_value threshold, a tunable parameter that determines which relationships are included in the graph.

    \item Attention Mechanism: The attention mechanism calculates the importance of each stock’s neighbors using the following equation:
          \begin{equation}
              \centering
              \begin{split}
                  e_{ij}^{'}=LeakyReLU(a'^{\top})[W_{z}Z_{i}||W_{z}Z_{j}])
              \end{split}
              \centering
          \end{equation}
          where $Z_{i} \in \mathbb{R}^{d_{z}}$ and $Z_{j} \in \mathbb{R}^{d_{z}}$ are the concatenated feature vectors of stock $i$ and stock $j$, $W_{z} \in \mathbb{R}^{d_{z'} \times d_{z}}$ is a learnable weight matrix, $a'$ is a learnable attention vector, $||$ represents concatenation, $d_{z'}$ is the size of the dimension.. The attention scores are then normalized across all neighboring stocks using a softmax function:
          \begin{equation}
              \centering
              \begin{split}
                  \sigma_{ij}^{'} = \frac{\exp(e_{ij}^{'})}{\sum_{k \in N'(i)}^{}exp(e_{ik}^{'})}
              \end{split}
              \centering
          \end{equation}
          where $N'_{i}$ denotes the set of neighbors of the node $i$.

    \item Feature Aggregation: The final output for each stock is computed as a weighted sum of its neighbors' features, with the attention weights $\sigma_{ij}^{'}$ determining the contribution of each neighbor:
          \begin{equation}
              \centering
              \begin{split}
                  Z^{'}_{i} = \sum_{j \in N'(i)}^{}\sigma_{ij}^{'}W_{z}Z_{j}
              \end{split}
              \centering
          \end{equation}
          where $Z^{'}_{i} \in \mathbb{R}^{d_{z'}}$ effectively integrates information from neighboring stocks, leading to a refined feature representation that incorporates both temporal, cross-sectional, and latent market features.

    \item Dimensionality Reduction: Put the output $Z^{'}_{i}$ of the first GAT layer into the second GAT layer to reduce the dimension. The calculation method is the same as above, and the final output $Z^{''}_{i}$ is obtained as the prediction result of each stock.
\end{itemize}

Although both GATs are used, their roles differ. The first GAT focuses on extracting cross-sectional information, capturing static dependencies between stocks based on historical correlations, and generating $A_{2}$. The second GAT, with input $Z$ (a fusion of multiple features), handles multimodal information fusion, captures higher-order relationships, and refines features. It combines information from different sources, learns complex correlations like indirect and dynamic relationships, and performs dimensionality reduction for more compact and efficient feature representations for final prediction.

\subsubsection{Loss Function}
After the final prediction is obtained from the GAT layers, the next step is to compute the loss, which measures the difference between the model’s predicted stock returns and the actual values. The choice of the loss function is critical, as it guides the model’s training process and influences its performance. In this model, we use mean squared error (MSE) for stock prediction tasks.  It is defined as:
\begin{equation}
    \centering
    \begin{split}
        Loss = \frac{1}{N}\sum_{i=1}^{N}(Z^{''}_{i}-y_{i})^{2}
    \end{split}
    \centering
\end{equation}
where $Z^{''}_{i}$ is the predicted value for the stock $i$ and $y_{i}$ is the actual value. During training, the model parameters are optimized to minimize the chosen loss function, leading to better predictive performance over time. We use Adam as a gradient-based optimization algorithm to update the model’s parameters.

\begin{table*}[t]
    \centering
    \caption{Details of the CSI 300, CSI 500, NASDAQ 100, and S\&P 500 dataset.}
    \label{table1}
    \resizebox{\textwidth}{!}{
        \begin{tabular}{c|cc}
            \toprule
            Stock Market             & CSI300                                    & CSI500                                          \\
            \midrule
            Number of Stocks         & 285                                       & 450                                             \\
            Year Established         & 2005                                      & 2007                                            \\
            Industry Coverage        & Broad Coverage                            & Broad Coverage                                  \\
            Market Cap Range         & Large Cap                                 & Mid \& Small Cap                                \\
            Representative Companies & ICBC, Vanke                               & Citic Bank, Shuanghui                           \\
            Market                   & China A-Share                             & China A-Share                                   \\
            Calculation Method       & Free-float Market Cap Weighted            & Free-float Market Cap Weighted                  \\
            Coverage Ratio           & Covers 70\% of China's A-share market cap & Covers 85\% of China's A-share market cap       \\
            \midrule
            Stock Market             & NASDAQ100                                 & S\&P500                                         \\
            \midrule
            Number of Stocks         & 99                                        & 498                                             \\
            Year Established         & 1985                                      & 1957                                            \\
            Industry Coverage        & Primarily Tech                            & Broad Coverage                                  \\
            Market Cap Range         & Large Cap                                 & Large Cap                                       \\
            Representative Companies & Apple, Microsoft                          & Apple, Amazon                                   \\
            Market                   & US Stock Market                           & US Stock Market                                 \\
            Calculation Method       & Free-float Market Cap Weighted            & Free-float Market Cap Weighted                  \\
            Coverage Ratio           & Covers 90\% of NASDAQ's total market cap  & Covers 80\% of NYSE and NASDAQ total market cap \\
            \bottomrule
        \end{tabular}}
\end{table*}

\section{Experiments}
In this section, we will thoroughly discuss the experimental design, including the experimental setup, baseline models, results, parameter sensitivity analysis, ablation studies, and case analysis. The experimental setup section specifically covers the datasets, evaluation metrics, and model parameters.

\subsection{Experimental Setttings}

\subsubsection{\textbf{Datasets}}
We employ four distinct stock market datasets to rigorously assess the robustness and generalizability of our model across varying market conditions, the details of the dataset are shown in Table \ref{table1}. Our selection encompasses the Shanghai-Shenzhen CSI 300\footnote{https://cn.investing.com/indices/csi300} and CSI 500\footnote{https://cn.investing.com/indices/china-securities-500-historical-data} datasets, which provide comprehensive coverage of the Chinese market's large-cap and mid-cap sectors, respectively. These datasets enable a detailed exploration of the dynamics within one of the world's largest and most complex financial markets. In contrast, the S\&P 500\footnote{https://hk.finance.yahoo.com/quote/\%5EGSPC/history/} dataset represents 500 leading companies across the U.S. market, offering insights into a broad and diverse economic landscape. Additionally, the NASDAQ 100\footnote{https://hk.finance.yahoo.com/quote/\%5EIXIC/history} dataset highlights the top 100 non-financial firms listed on the NASDAQ, with a particular focus on the technology sector's rapidly evolving dynamics. Collectively, these datasets provide a comprehensive view of different market behaviors and geographic regions, thereby supporting a robust evaluation of our model's predictive capabilities across varied financial contexts.

We structure our datasets following a temporal sequence, dividing them into distinct phases for training (from January 1, 2018, to December 31, 2021), validation (from January 1, 2022, to December 31, 2022), and testing (from January 1, 2023, to December 31, 2023). In our forecasting approach, we employ features derived from the previous 60 trading days to predict stock return rankings over the next 21 trading days. This methodology closely mirrors the decision-making process in real-world trading scenarios. For the baseline analysis, we utilize data from the four aforementioned stock markets, concentrating on six key financial indicators: open price, close price, high price, low price, turnover, and volume. We commence by implementing procedures for outlier detection and normalization to ensure data integrity and reduce the impact of anomalous values. Following this, we calculate the daily return for each stock as the label during training, defined as the percentage change between the closing prices of consecutive trading days.

\subsubsection{\textbf{Evaluation Metrics}}
The trading strategy simulated by our model is outlined as follows:
\begin{itemize}
    \item At the close of trading day $t$, the model generates a prediction score for each stock, ranking them based on the expected rate of return.

    \item At the opening of trading day $t+1$, traders liquidate the stocks purchased on day $t$ and acquire those ranked in the top-$k$ for expected returns.

    \item If a stock consistently ranks among the highest expected returns, it remains in the trader's portfolio.

    \item Transaction costs are excluded from consideration in this simulation.
\end{itemize}
In order to improve the reliability of the evaluation, we conduct ten training and predictions for each method and take the average result of the ten times as the final prediction result, and then trade the strategy.

The primary objective is to identify stocks with the highest returns and to evaluate the performance of both the baseline and our proposed model, we employ six key financial metrics:
\begin{itemize}
    \item Annualized Rate of Return~(ARR): This core metric aggregates the daily returns of selected stocks over a year, indicating the effectiveness of the investment strategy. It is computed as $ARR=(\prod_{T}^{t+1}(1+r_{t}))^{\frac{252}{T}}-1$, where $r_{t}$ is the daily return and $T$ is the total number of trading days in the year.

    \item Annualized Volatility~(AVoL): This metric captures the annualized standard deviation of daily returns, representing the risk associated with the strategy. It is calculated as $AVoL = std(\frac{P_{t}}{P_{t-1}} -1)*\sqrt{252}$, where $P_{t}$ and $P_{t-1}$ are the stock prices on day $t$ and day $t-1$, respectively.

    \item Maximum Drawdown~(MDD): MDD measures the most substantial decline from a peak to a trough during the testing period, indicating the strategy's potential risk of loss. It is computed as $\underset{t\in[1,T]}{max}(\frac{max_{i \in [1, t]}P_{i} -P_{t}}{max_{i \in [1, t]}P_{i} })$, where $P_{t}$ is the price of the stock at time $t$, and $T$ is the length of the period.

    \item Annualized Sharpe Ratio~(ASR): This metric evaluates the return per unit of volatility and is calculated as $ASR = \frac{ARR}{AVoL}$, reflecting the risk-adjusted performance of the strategy.

    \item Calmar Ratio~(CR): CR assesses the return relative to the maximum drawdown, calculated as $CR = \frac{ARR}{|MDD|}$, offering insight into the return-risk trade-off.

    \item Information Ratio~(IR): This metric measures the excess return per unit of additional risk, further refining the assessment of the strategy’s risk-adjusted performance. It is calculated as $IR = \frac{mean(r_{t}-r_{f,t})}{std(r_{t}-r_{f,t})}$, where $r_{t}$ is the portfolio return at the time $t$
          and $r_{f,t}$ is the return of a benchmark or risk-free asset at the same time.

\end{itemize}
Together, these metrics form a comprehensive framework for evaluating both the performance and risk profile of the investment strategies. Higher values of ARR, ASR, CR, and IR, coupled with lower values of AVoL and MDD, signify superior performance.

\begin{table*}[t]
    \centering
    \caption{Experimental parameter summary table.}
    \label{table1_2}
    \resizebox{\textwidth}{!}{
        \begin{tabular}{c|c|c} \toprule
            \multicolumn{1}{c|}{Module Name} & \multicolumn{1}{c|}{Parameter Name}                  & \multicolumn{1}{c}{Value} \\ \hline
            \multirow{3}{*}{Use Improved GRU to Capture Temporal Features}
                                             & number of GRU layers                                 & 2                         \\
                                             & number of neurons in the first layer                 & 32                        \\
                                             & number of neurons in the second layer                & 10                        \\ \hline
            \multirow{5}{*}{Use GAT to Capture Cross-sectional Features}
                                             & number of GAT layers                                 & 2                         \\
                                             & number of neurons in the first layer                 & 32                        \\
                                             & number of neurons in the second layer                & 4                         \\
                                             & number of attention heads                            & 4                         \\
                                             & threshold for connecting stock nodes $judge_{value}$ & 0.8                       \\ \hline
            \multirow{2}{*}{Use Multi-head Cross-attention to Capture Latent State Feature}
                                             & number of latent states $d_r$                        & 32                        \\
                                             & dimension of each latent state $d_i$                 & 16                        \\ \hline
            \multirow{2}{*}{Model Prediction and Loss Calculation Layer}
                                             & number of GAT layers                                 & 1                         \\
                                             & number of neurons in the first layer                 & 32                        \\ \hline
            \multirow{4}{*}{Training Process}
                                             & batch size                                           & 32                        \\
                                             & loss  function                                       & MSE                       \\
                                             & optimizer                                            & adam                      \\
                                             & learning rate                                        & 0.0002                    \\ \bottomrule
        \end{tabular}
    }
\end{table*}

\subsubsection{\textbf{Parameter Settings}}
We configure the time window $t$ to include the historical stock data from the previous 10 days as the model's input for training. The model architecture consists of four primary components, each designed to capture different patterns in financial time series. Each of these components contains its own set of parameters, as summarized in Table \ref{table1_2}.

In the ``Use Improved GRU to Capture Temporal Features'' module, we incorporate an attention mechanism to replace the traditional reset gate in the GRU framework. This module consists of two layers: the first layer contains 32 neurons, and the second layer contains 10 neurons. The output of this module is represented as $A_{1}$, which is responsible for extracting temporal dependencies from the stock data within the historical window.
In the ``Use GAT to Capture Cross-sectional Features'' module, the architecture consists of two layers, each with tunable parameters. The number of attention heads is a key factor affecting the model's expressive capability, and it was fine-tuned through experiments. We tested four values: 2, 4, 6, and 8, and ultimately chose 4. The initial GAT layer consists of 32 neurons and 4 attention heads, while the second GAT layer reduces the feature dimension to 4 neurons. The relationships between the stock nodes in the graph are defined based on the correlation of stock returns from the previous year. During training, a threshold value of $judge_{value} = 0.8$ is used to filter stock pairs with significant correlations. The $judge_{value}$ parameter was optimized through experimental tuning, testing values of 0.2, 0.4, 0.6, 0.8, and 1.0, with 0.8 being the final choice. The output of the GAT layer is represented as $A_{2}$, which captures the structural dependencies inherent in the stock correlation graph.
In the ``Use Multi-head Cross-attention to Capture Latent State Feature'' module, two learnable market latent state vectors, $ R_{1} \in \mathbb{R}^{d_{h} \times d_{i}} $ and $R_{2} \in \mathbb{R}^{d_{h} \times d_{i}}$, are initialized, both with dimensions $(32, 16)$. The choice of the hyperparameters $d_r$ (number of latent states) and $d_i$(dimension of each latent state) is typically influenced by the complexity of the data and the task requirements. We optimized these parameters through grid search to balance computational efficiency and model expressiveness. We tested $d_r$ values of 8, 16, 24, 32, and 40, ultimately selecting 32. For $d_i$, we tested values of 4, 8, 12, 16, and 20, with 16 being chosen as the final value. The multi-head cross-attention mechanism involves interactions between $R_{1}$ and $A_{1}$, as well as $R_{2}$ and $A_{2}$, producing representation vectors $B_{1}$ and $B_{2}$. With 4 attention heads, the multi-head cross-attention mechanism encapsulates latent market states by leveraging temporal and structural features derived from the stock data.
Finally, the ```Model Prediction and Loss Calculation Layer'' integrates the outputs $A_{1}$, $A_{2}$, $B_{1}$, and $B_{2}$. These outputs are concatenated and passed through one additional layers of GAT for final prediction. This final module is responsible for predicting future stock trends by merging both temporal and relational information.

The batch size of 32 is chosen based on experimental tuning. While 32 is a commonly used value and works well for most models, we acknowledge that it can influence the stability and efficiency of the training process. We selected 32 as a compromise, ensuring efficient utilization of computational resources while avoiding excessively large variance during training. Future research may further explore different batch sizes through experiments to optimize performance for specific tasks and datasets.
For the loss function, we used Mean Squared Error (MSE), which is the most common choice for regression tasks, particularly when the target variable is continuous. We opted for MSE because it provides stable performance in most regression tasks and effectively measures the squared error of predictions.
The Adam optimizer was chosen due to its strong performance across various deep learning tasks, particularly its adaptability in the optimization process, which allows for good convergence with minimal tuning. The initial learning rate of 0.0002 was determined through preliminary experiments, with smaller learning rates typically ensuring stable convergence. This value was selected based on its performance in our experiments but can be adjusted depending on the specific task at hand. For practical application, each trading day a virtual investment portfolio is constructed, comprising the top 10 stocks ranked by the model's predicted returns.

\subsection{Baseline Models}
We conduct a comparative analysis between our proposed MCI-GRU model and a range of baseline models, which include prominent approaches in both traditional machine learning and deep learning for time series prediction, as well as reinforcement learning for portfolio management.
\begin{itemize}
    \item \textbf{BLSW} \cite{BLSW}: Implements a mean reversion trading strategy, which assumes that asset prices will revert to their historical average over time, making it particularly effective in markets with cyclical behavior.

    \item \textbf{Cross-Sectional Mean Reversion (CSM)} \cite{CSM}: Adopts a momentum-based approach by identifying assets that exhibit persistent price trends, and positioning trades in alignment with these trends, thereby capitalizing on short-term market movements.

    \item \textbf{LSTM} \cite{LSTM}: A widely used recurrent neural network model for time series forecasting, which captures temporal dependencies through its memory cell mechanism.

    \item \textbf{ALSTM} \cite{ALSTM}: An enhanced version of LSTM, which incorporates dual attention mechanisms, one for adaptively selecting relevant features and another for focusing on significant time steps, thereby improving prediction accuracy by concentrating on key information.

    \item \textbf{GRU} \cite{GRU}: A simplified variant of LSTM, using fewer gating mechanisms to streamline the learning process and improve computational efficiency while maintaining strong performance in sequence prediction tasks.

    \item \textbf{Transformer} \cite{Trans, Transformer}: Utilizes a multi-head self-attention mechanism to capture long-range dependencies in time series data, with the ability to process entire sequences in parallel, offering scalability and improved performance over recurrent architectures.

    \item \textbf{TRA} \cite{TRA}: Introduces a novel dynamic routing mechanism within the Transformer architecture, enabling the model to adaptively learn temporal patterns in stock prices, improving its ability to capture diverse market trends.

    \item \textbf{CTTS} \cite{CTTS}: Combines convolutional neural networks (CNNs) with Transformer layers to capture both local feature patterns and global temporal dependencies in financial data, providing a hybrid approach to time series forecasting.

    \item \textbf{A2C} \cite{A2C}: A deep reinforcement learning method employing parallel actor-learners and asynchronous gradient descent for policy optimization, facilitating efficient exploration and exploitation in large action spaces.

    \item \textbf{DDPG} \cite{DDPG}: A deterministic deep reinforcement learning algorithm that extends DQN by incorporating policy gradients, designed specifically for continuous action spaces, leveraging an off-policy actor-critic architecture for stable learning.

    \item \textbf{PPO} \cite{PPO}: Optimizes policies using a clipped surrogate objective, balancing policy exploration and stability through mini-batch updates, making it robust in volatile market environments.

    \item \textbf{TD3} \cite{TD3}: Builds on DDPG by introducing three key innovations: twin Critic networks to reduce overestimation bias, delayed updates to the Actor-network for stability, and adding noise to the policy during training to improve exploration.

    \item \textbf{SAC} \cite{SAC}: An off-policy deep reinforcement learning approach that incorporates entropy regularization to encourage exploration, optimizing a stochastic policy for continuous actions, with dual Critic networks to improve value estimation accuracy.

    \item \textbf{FactorVAE} \cite{FactorVAE}: Integrates dynamic factor models with a variational autoencoder, enabling the prediction of cross-sectional stock returns by modeling latent factors that drive asset price movements.

    \item \textbf{AlphaStock} \cite{AlphaStock}: A hybrid model that combines deep learning and reinforcement learning with a cross-asset attention mechanism to capture the intricate relationships between assets, enhancing stock prediction accuracy by exploiting interdependencies.

    \item \textbf{DeepPocket} \cite{DeepPocket}: Merges graph neural networks, autoencoders, and reinforcement learning in a unified framework, focusing on managing financial portfolios by modeling the latent relationships between assets for dynamic decision-making.

    \item \textbf{DeepTrader} \cite{DeepTrader}: Utilizes deep reinforcement learning with graph convolutional networks to model the interrelationships between stocks, leveraging industry classifications and causal dependencies to capture both spatial and temporal market dynamics for effective portfolio management.

    \item \textbf{THGNN} \cite{THGNN}: A sophisticated temporal-heterogeneous graph neural network that integrates dynamic company relationships with Transformer encoders, featuring a two-stage attention mechanism to enhance financial time series prediction by focusing on critical temporal and structural patterns.

    \item \textbf{VGNN} \cite{xing2023learning}: A decoupled graph learning framework for vague graph learning in stock prediction, leveraging matrix/tensor fusion, hybrid attention, and message passing to outperform state-of-the-art methods.

    \item \textbf{MASTER} \cite{MASTER}: A market-centric transformer model that dynamically simulates both instantaneous and cross-temporal correlations between stocks, enhancing trend forecast precision.

    \item \textbf{CISTHPAN} \cite{xia2024ci}: A two-stage framework combining Transformer and HGAT-based pre-training with stock-ranking fine-tuning, designed to improve the accuracy of stock movement predictions.

    \item \textbf{CL} \cite{du2024explainable}: A model combining textual and quantitative indicators for improved explainability in stock movement prediction, achieving state-of-the-art performance.
\end{itemize}

\begin{table*}[t]
    \centering
    \caption{Comparing the experimental results of the models on CSI 300 and CSI 500 datasets. ARR measures the portfolio return rate of each predictive model, with higher values being better. AVol and MDD measure the investment risk of each predictive model, with lower absolute values being better. ASR, CR, and IR measure profits under unit risk, with higher values being better.}
    \label{table2}
    \resizebox{\textwidth}{!}{
        \begin{tabular}{c|cccccccc|cccccccc}
            \toprule
            Datasets         & \multicolumn{8}{c|}{CSI 300} & \multicolumn{8}{c}{CSI 500}                                                                                                                                                                                                                                                      \\
            Model            & ARR $\uparrow$               & AVol $\downarrow$           & MDD $\downarrow$ & ASR $\uparrow$ & CR $\uparrow$  & IR $\uparrow$  & MSE $\downarrow$ & MAE $\downarrow$
                             & ARR $\uparrow$               & AVol $\downarrow$           & MDD $\downarrow$ & ASR $\uparrow$ & CR $\uparrow$  & IR $\uparrow$  & MSE $\downarrow$ & MAE $\downarrow$                                                                                                                                          \\
            \midrule
            BLSW             & -0.076                       & 0.113                       & -0.231           & -0.670         & -0.316         & 0.311          & 0.078            & 0.090            & 0.110          & 0.227          & -0.155          & 0.485          & 0.710          & 0.446          & 0.033          & 0.064          \\
            CSM              & -0.185                       & 0.204                       & -0.293           & -0.907         & -0.631         & -0.935         & 0.093            & 0.124            & 0.015          & 0.229          & -0.179          & 0.066          & 0.084          & 0.001          & 0.049          & 0.098          \\
            LSTM             & -0.214                       & 0.175                       & -0.275           & -1.361         & -0.779         & -1.492         & 0.108            & 0.142            & -0.008         & 0.159          & -0.172          & -0.047         & -0.044         & -0.128         & 0.065          & 0.118          \\
            ALSTM            & -0.216                       & 0.164                       & -0.294           & -1.314         & -0.735         & -1.461         & 0.112            & 0.148            & 0.016          & 0.162          & -0.192          & 0.101          & 0.086          & 0.014          & 0.045          & 0.093          \\
            GRU              & -0.229                       & 0.156                       & -0.290           & -1.469         & -0.790         & -1.631         & 0.123            & 0.155            & -0.004         & 0.159          & -0.193          & -0.028         & -0.023         & -0.118         & 0.062          & 0.112          \\
            Transformer      & -0.240                       & 0.156                       & -0.281           & -1.543         & -0.855         & -1.695         & 0.131            & 0.159            & 0.154          & 0.156          & -0.135          & 0.986          & 1.143          & 0.867          & 0.025          & 0.053          \\
            \hline
            TRA              & -0.074                       & 0.169                       & -0.222           & -0.436         & -0.332         & -0.409         & 0.079            & 0.088            & 0.125          & 0.162          & -0.145          & 0.776          & 0.866          & 0.657          & 0.030          & 0.062          \\
            CTTS             & -0.193                       & 0.206                       & -0.312           & -0.937         & -0.618         & -0.907         & 0.095            & 0.129            & -0.041         & 0.172          & -0.239          & -0.241         & -0.173         & -0.237         & 0.078          & 0.131          \\
            A2C              & -0.207                       & 0.092                       & -0.259           & -2.255         & -0.803         & -2.490         & 0.099            & 0.137            & -0.172         & 0.084          & -0.208          & -2.043         & -0.826         & -2.207         & 0.098          & 0.166          \\
            DDPG             & -0.137                       & 0.138                       & -0.240           & -0.992         & -0.568         & -1.002         & 0.087            & 0.106            & -0.128         & 0.082          & -0.170          & -1.563         & -0.756         & -1.639         & 0.088          & 0.145          \\
            PPO              & -0.096                       & \textbf{0.045}              & \textbf{-0.120}  & -2.138         & -0.800         & -2.234         & 0.083            & 0.096            & -0.032         & \textbf{0.015} & \textbf{-0.040} & -2.041         & -0.787         & -2.075         & 0.073          & 0.127          \\
            TD3              & -0.154                       & 0.137                       & -0.252           & -1.122         & -0.610         & -1.155         & 0.090            & 0.115            & -0.123         & 0.135          & -0.248          & -0.912         & -0.496         & -0.909         & 0.085          & 0.139          \\
            SAC              & -0.140                       & 0.090                       & -0.207           & -1.554         & -0.676         & -1.635         & 0.086            & 0.109            & -0.167         & 0.081          & -0.207          & -2.057         & -0.807         & -2.219         & 0.095          & 0.158          \\
            \hline
            FactorVAE        & -0.048                       & 0.134                       & -0.175           & -0.335         & -0.271         & -0.348         & 0.071            & 0.085            & 0.006          & 0.127          & -0.147          & 0.047          & 0.041          & 0.112          & 0.055          & 0.103          \\
            AlphaStock       & -0.164                       & 0.153                       & -0.245           & -1.072         & -0.669         & -1.098         & 0.092            & 0.112            & -0.017         & 0.148          & -0.166          & 0.115          & -0.102         & -0.043         & 0.069          & 0.122          \\
            DeepPocket       & -0.036                       & 0.135                       & -0.175           & -0.270         & -0.207         & -0.258         & 0.065            & 0.080            & 0.006          & 0.127          & -0.148          & 0.050          & 0.043          & 0.115          & 0.056          & 0.104          \\
            DeepTrader       & -0.122                       & 0.147                       & -0.229           & -0.828         & -0.533         & -0.876         & 0.084            & 0.102            & 0.055          & 0.168          & -0.141          & 0.324          & 0.388          & 0.370          & 0.039          & 0.072          \\
            THGNN            & -0.015                       & 0.172                       & -0.152           & -0.088         & -0.100         & -0.003         & 0.043            & 0.071            & 0.048          & 0.128          & -0.141          & 0.375          & 0.340          & 0.432          & 0.045          & 0.078          \\
            VGNN             & -0.037                       & 0.163                       & -0.197           & -0.227         & -0.188         & -0.201         & 0.061            & 0.082            & 0.111          & 0.166          & -0.175          & 0.668          & 0.636          & 0.564          & 0.031          & 0.063          \\
            MASTER           & 0.102                        & 0.151                       & -0.126           & 0.681          & 0.811          & 0.726          & 0.057            & 0.078            & 0.128          & 0.130          & -0.098          & 0.989          & 1.308          & 0.997          & 0.028          & 0.057          \\
            CI-STHPAN        & -0.078                       & 0.167                       & -0.144           & -0.466         & -0.538         & -0.355         & 0.075            & 0.092            & 0.021          & 0.151          & -0.129          & 0.136          & 0.161          & 0.211          & 0.053          & 0.088          \\
            CL               & -0.035                       & 0.148                       & -0.183           & -0.241         & -0.195         & -0.193         & 0.063            & 0.078            & 0.051          & 0.146          & -0.128          & 0.351          & 0.401          & 0.390          & 0.041          & 0.075          \\
            \hline
            \textbf{MCI-GRU} & \textbf{0.352}               & 0.226                       & -0.127           & \textbf{1.559} & \textbf{2.776} & \textbf{1.526} & \textbf{0.035}   & \textbf{0.068}   & \textbf{0.330} & 0.203          & -0.198          & \textbf{1.626} & \textbf{1.663} & \textbf{1.382} & \textbf{0.022} & \textbf{0.051} \\
            \bottomrule
        \end{tabular}}
\end{table*}

\begin{table*}[t]
    \centering
    \caption{Comparing the experimental results of the models on S\&P 500 and NASDAQ 100 datasets.}
    \label{table3}
    \resizebox{\textwidth}{!}{
        \begin{tabular}{c|cccccccc|cccccccc}
            \toprule
            Datasets    & \multicolumn{8}{c|}{S\&P 500} & \multicolumn{8}{c}{NASDAQ 100}                                                                                                                                                                                                \\
            Model       & ARR $\uparrow$                & AVol $\downarrow$              & MDD $\downarrow$ & ASR $\uparrow$ & CR $\uparrow$  & IR $\uparrow$  & MSE $\downarrow$ & MAE $\downarrow$
                        & ARR $\uparrow$                & AVol $\downarrow$              & MDD $\downarrow$ & ASR $\uparrow$ & CR $\uparrow$  & IR $\uparrow$  & MSE $\downarrow$ & MAE $\downarrow$                                                                                    \\
            \midrule
            BLSW        & 0.199                         & 0.318                          & -0.223           & 0.626          & 0.892          & 0.774          & 0.068            & 0.094            & 0.368 & 0.339          & -0.222          & 1.086 & 1.658 & 1.194 & 0.058 & 0.082 \\
            CSM         & 0.099                         & 0.250                          & -0.139           & 0.396          & 0.712          & 0.584          & 0.125            & 0.173            & 0.116 & 0.242          & -0.145          & 0.479 & 0.800 & 0.603 & 0.119 & 0.178 \\
            LSTM        & 0.142                         & 0.162                          & -0.178           & 0.877          & 0.798          & 0.929          & 0.097            & 0.132            & 0.247 & 0.176          & -0.128          & 1.403 & 1.930 & 1.386 & 0.085 & 0.129 \\
            ALSTM       & 0.191                         & 0.161                          & -0.150           & 1.186          & 1.273          & 1.115          & 0.073            & 0.098            & 0.201 & 0.192          & -0.183          & 1.047 & 1.098 & 1.032 & 0.084 & 0.147 \\
            GRU         & 0.124                         & 0.169                          & -0.139           & 0.734          & 0.829          & 1.023          & 0.109            & 0.151            & 0.225 & 0.188          & -0.165          & 1.197 & 1.364 & 1.160 & 0.089 & 0.138 \\
            Transformer & 0.135                         & 0.159                          & -0.140           & 0.852          & 0.968          & 0.908          & 0.103            & 0.145            & 0.268 & 0.175          & -0.131          & 1.531 & 2.046 & 1.441 & 0.082 & 0.121 \\
            \hline
            TRA         & 0.184                         & 0.166                          & -0.158           & 1.114          & 1.172          & 1.106          & 0.079            & 0.104            & 0.267 & 0.181          & -0.144          & 1.475 & 1.854 & 1.427 & 0.081 & 0.119 \\
            CTTS        & 0.154                         & 0.161                          & -0.113           & 0.952          & 1.356          & 0.965          & 0.093            & 0.125            & 0.349 & 0.197          & -0.193          & 1.769 & 1.808 & 1.610 & 0.071 & 0.102 \\
            A2C         & 0.160                         & 0.126                          & -0.084           & 1.267          & 1.907          & 1.244          & 0.089            & 0.117            & 0.109 & 0.134          & -0.114          & 0.816 & 0.957 & 0.844 & 0.126 & 0.182 \\
            DDPG        & 0.111                         & 0.129                          & -0.091           & 0.864          & 1.223          & 0.887          & 0.121            & 0.165            & 0.130 & 0.156          & -0.131          & 0.832 & 0.994 & 0.863 & 0.110 & 0.176 \\
            PPO         & 0.020                         & \textbf{0.089}                 & \textbf{-0.067}  & 0.220          & 0.291          & 0.263          & 0.159            & 0.212            & 0.148 & \textbf{0.118} & \textbf{-0.104} & 1.259 & 1.424 & 1.237 & 0.104 & 0.165 \\
            TD3         & 0.024                         & 0.113                          & -0.105           & 0.209          & 0.225          & 0.264          & 0.151            & 0.201            & 0.181 & 0.155          & -0.160          & 1.169 & 1.130 & 1.156 & 0.093 & 0.160 \\
            SAC         & 0.140                         & 0.111                          & -0.069           & 1.263          & 2.011          & 1.242          & 0.099            & 0.136            & 0.162 & 0.139          & -0.107          & 1.165 & 1.518 & 1.154 & 0.098 & 0.158 \\
            \hline
            FactorVAE   & 0.160                         & 0.142                          & -0.132           & 1.128          & 1.211          & 1.013          & 0.088            & 0.115            & 0.356 & 0.159          & -0.119          & 2.234 & 2.995 & 1.907 & 0.063 & 0.089 \\
            AlphaStock  & 0.122                         & 0.140                          & -0.126           & 0.871          & 0.968          & 0.892          & 0.112            & 0.157            & 0.372 & 0.178          & -0.134          & 1.781 & 2.776 & 1.869 & 0.053 & 0.084 \\
            DeepPocket  & 0.165                         & 0.142                          & -0.126           & 1.165          & 1.311          & 1.045          & 0.084            & 0.112            & 0.346 & 0.157          & -0.116          & 2.197 & 2.971 & 1.882 & 0.061 & 0.098 \\
            DeepTrader  & 0.295                         & 0.180                          & -0.181           & 1.635          & 1.628          & 1.425          & 0.049            & 0.082            & 0.716 & 0.248          & -0.138          & 2.890 & 5.169 & 2.306 & 0.019 & 0.048 \\
            THGNN       & 0.271                         & 0.141                          & -0.094           & 1.921          & 2.871          & 1.778          & 0.057            & 0.086            & 0.644 & 0.204          & -0.146          & 3.147 & 3.414 & 2.543 & 0.026 & 0.057 \\
            VGNN        & 0.299                         & 0.202                          & -0.169           & 1.473          & 1.767          & 1.406          & 0.045            & 0.076            & 0.616 & 0.181          & -0.091          & 3.405 & 6.767 & 2.798 & 0.033 & 0.063 \\
            MASTER      & 0.335                         & 0.171                          & -0.134           & 1.958          & 2.495          & 1.895          & 0.038            & 0.065            & 0.654 & 0.168          & -0.076          & 3.876 & 8.534 & 3.183 & 0.023 & 0.052 \\
            CI-STHPAN   & 0.123                         & 0.233                          & -0.254           & 0.527          & 0.485          & 0.632          & 0.114            & 0.159            & 0.454 & 0.208          & -0.124          & 2.178 & 3.660 & 1.855 & 0.041 & 0.071 \\
            CL          & 0.308                         & 0.189                          & -0.171           & 1.629          & 1.794          & 1.451          & 0.039            & 0.071            & 0.351 & 0.172          & -0.122          & 2.041 & 2.877 & 1.821 & 0.066 & 0.093 \\
            \textbf{MCI-GRU}
                        & \textbf{0.456}                & 0.179                          & -0.129           & \textbf{2.549} & \textbf{3.543} & \textbf{2.197} & \textbf{0.031}   & \textbf{0.062}
                        & \textbf{0.718}                & 0.220                          & -0.118           & \textbf{3.257} & \textbf{6.091} & \textbf{2.609} & \textbf{0.015}   & \textbf{0.039}                                                                                      \\
            \bottomrule
        \end{tabular}}
\end{table*}

\subsection{Experimental Results}
In this section, we conduct a rigorous evaluation of the experimental results of our proposed model in comparison with several baseline models across different datasets, as illustrated in Tables \ref{table2} and \ref{table3}.

On the CSI 300 dataset, traditional and deep learning models (including BLSW, CSM, LSTM, ALSTM, GRU, and Transformer) generally exhibit subpar performance, characterized by negative ARR values and low ASR, CR, and IR metrics. For example, although the Transformer model achieves the highest ARR among the traditional models, its ARR is still -0.240, with an MDD of -0.281, indicating significant risk exposure. In the case of the CSI 500 dataset, the performance shows slight improvement, with the Transformer model achieving an ARR of 0.154 and an ASR of 0.986. However, these values are still significantly lower than those offered by our proposed model. While reinforcement learning models (such as TRA, CTTS, A2C, DDPG, PPO, TD3, SAC, and FactorVAE) show marginal improvements, their performance remains below that of our model. For instance, the PPO model has the lowest AVoL and MDD, indicating lower risk, but its ARR and IR metrics are still negative in both datasets, highlighting poor return potential. Graph-based models (including AlphaStock, DeepPocket, DeepTrader, THGNN, VGNN, MASTER, and CISTHPAN) show more promising results. For example, THGNN achieves an ARR of -0.015 on the CSI 300 and an ARR of 0.048 on the CSI 500. Among these graph-based models, VGNN demonstrates relatively good ARR values on both datasets, with -0.037 on the CSI 300 and 0.111 on the CSI 500, but its risk-adjusted return metrics (ASR, CR, IR) still require improvement. The MASTER model achieves a relatively high ARR of 0.102 on the CSI 300, along with relatively high ASR, CR, and IR, showing some potential, but its performance on the CSI 500 is comparatively mediocre. The CI-STHPAN model performs poorly on both datasets, with low ARR and high-risk metrics. The CL model performs poorly on the CSI 300 but shows some improvement on the CSI 500 with an ARR of 0.051, though its volatility is also relatively high. Despite their individual characteristics and performances, the performance of these graph-based models still falls short compared to our model.

Overall, the model performs better in the U.S. market than in the Chinese market based on the ARR metric. In particular, the NASDAQ 100 index shows high ARR values, with MCI-GRU reaching 0.718, reflecting strong growth in the U.S. tech market. In contrast, ARR values in the Chinese market are generally lower, and many traditional models yield negative ARR for the CSI 300 index. This discrepancy may stem from higher volatility, a more complex market structure, and investor behavior in China. Even the best-performing MCI-GRU achieves only 0.352 on the CSI 300 index, lower than in the U.S.
The U.S. market exhibits relatively low volatility, particularly in the NASDAQ 100, where many models have aAVol below 0.2, indicating stability. Conversely, the Chinese market shows higher volatility, likely due to its younger market environment and dominance of retail investors, which complicates prediction and risk management. Regarding MDD, the U.S. market typically shows smaller MDD values, reinforcing its stability, while the Chinese market experiences higher MDD, indicating greater downside risk. In terms of risk-adjusted metrics—ASR, CR, and IR—the U.S. market shows higher values, driven by higher ARR and lower volatility, suggesting better risk-adjusted returns. The Chinese market has lower risk-adjusted returns, primarily due to lower ARR and higher volatility. The MCI-GRU model performs well across all four datasets, demonstrating robustness and adaptability in different market environments. In contrast, traditional and reinforcement learning models perform notably worse in the Chinese market, indicating they may be better suited for more mature, stable markets.

Additionally, as shown in Tables \ref{table2} and \ref{table3}, there are significant differences in the performance of various models in terms of Mean Squared Error (MSE) and Mean Absolute Error (MAE). Since this study focuses primarily on predicting daily stock returns, and the prediction errors are relatively small, the resulting MSE and MAE values are correspondingly modest. Although MAE and MSE are commonly used metrics for evaluating prediction accuracy, there is some correlation between these metrics and investment return indicators (such as ARR, ASR, and IR). In general, models with higher accuracy are likely to yield better investment returns. However, it is important to note that lower MAE and MSE do not necessarily translate into higher investment returns, as returns are influenced by various factors, including trading strategies and risk control.
Among all the models, MCI-GRU performs exceptionally well in terms of MAE and MSE, particularly on the NASDAQ 100 dataset, where its MAE and MSE are the lowest, indicating a clear advantage in prediction accuracy. Its excellent performance in investment return metrics further supports the correlation between higher prediction accuracy and better investment returns. In contrast, some traditional and reinforcement learning models perform relatively poorly in terms of MAE and MSE, which is consistent with their performance in investment return metrics, suggesting that these models may not effectively capture the dynamic fluctuations of the market.
GNN-based models generally show better performance in terms of MAE and MSE, indicating that graph structural information helps improve prediction accuracy. Among these models, DeepTrader, THGNN, VGNN, and MASTER perform well on both MAE and MSE, further demonstrating the importance of graph structures in capturing complex relationships in the stock market and enhancing prediction accuracy.

Our proposed model demonstrates a significant performance advantage, achieving the highest ARR (0.352 for CSI 300 and 0.330 for CSI 500) and exhibiting superior risk-adjusted returns, as indicated by the highest ASR (1.559 for CSI 300 and 1.626 for CSI 500), CR (2.776 for CSI 300 and 1.663 for CSI 500), and IR (1.526 for CSI 300 and 1.382 for CSI 500). Similar trends are observed in Table \ref{table3}, which presents results from the S\&P 500 and NASDAQ 100 datasets. Traditional models such as BLSW and CSM show moderate performance, with CSM reporting an ARR of 0.099 and an ASR of 0.396 on the S\&P 500, while Transformer achieves an ARR of 0.135 and an ASR of 0.852. However, the inability of these models to effectively leverage relational data hampers their overall performance. Reinforcement learning models again demonstrate improvements, with SAC and FactorVAE achieving notable ASR values of 1.263 and 1.128, respectively, on the S\&P 500. On the NASDAQ 100, FactorVAE achieves an ARR of 0.356 and the highest ASR (2.234) among reinforcement learning models. Graph-based models, particularly DeepTrader and THGNN, excel on these datasets. For example, DeepTrader achieves an ARR of 0.716 and an ASR of 2.890 on the NASDAQ 100, while THGNN records an ARR of 0.644 and an ASR of 3.147, demonstrating the effectiveness of incorporating relational information into financial models. The superiority of our MCI-GRU model is further evident, as it consistently outperforms all baseline models. It achieves the highest ARR (0.456 for S\&P 500 and 0.718 for NASDAQ 100) and outstanding risk-adjusted returns, reflected in ASR values of 2.549 and 3.257, CR values of 3.543 and 6.091, and IR values of 2.197 and 2.609, respectively. These results underscore the model's effectiveness in capturing both long-term and short-term dependencies within financial data.

\begin{table*}[t]
    \centering
    \caption{The parameter sensitivity results of parameter judge\_value in dataset CSI 300, CSI 500, S\&P 500, and NASDAQ 100.}
    \label{table4}
    \resizebox{\textwidth}{!}{
        \begin{tabular}{c|cccccc|cccccc}
            \toprule
            Metrics  & ARR $\uparrow$                & AVol $\downarrow$              & MDD $\downarrow$ & ASR $\uparrow$ & CR $\uparrow$  & IR $\uparrow$
                     & ARR $\uparrow$                & AVol $\downarrow$              & MDD $\downarrow$ & ASR $\uparrow$ & CR $\uparrow$  & IR $\uparrow$  \\
            \midrule
            Datasets & \multicolumn{6}{c|}{CSI 300}  & \multicolumn{6}{c}{CSI 500}                                                                          \\
            \midrule
            \textbf{0.6}
                     & 0.102                         & 0.187                          & -0.142           & 0.546          & 0.720          & 0.650
                     & 0.107                         & \textbf{0.162}                 & -0.186           & 0.660          & 0.577          & 0.639          \\
            \textbf{0.7}
                     & 0.235                         & \textbf{0.183}                 & \textbf{-0.090}  & 1.282          & 2.623          & 1.201
                     & 0.212                         & 0.167                          & \textbf{-0.170}  & 1.267          & 1.245          & 1.217          \\
            \textbf{0.8}
                     & \textbf{0.352}                & 0.226                          & -0.127           & \textbf{1.559} & \textbf{2.776} & \textbf{1.526}
                     & \textbf{0.330}                & 0.203                          & -0.198           & \textbf{1.626} & \textbf{1.663} & \textbf{1.382} \\
            \textbf{0.9}
                     & 0.284                         & 0.241                          & -0.194           & 1.178          & 1.460          & 1.216
                     & 0.278                         & 0.205                          & -0.197           & 1.356          & 1.415          & 1.199          \\
            \midrule
            Datasets & \multicolumn{6}{c|}{S\&P 500} & \multicolumn{6}{c}{NASDAQ 100}                                                                       \\
            \midrule
            \textbf{0.6}
                     & 0.286                         & \textbf{0.162}                 & -0.138           & 1.768          & 2.069          & 1.611
                     & 0.532                         & \textbf{0.219}                 & -0.118           & 2.429          & 4.509          & 2.068          \\
            \textbf{0.7}
                     & 0.358                         & 0.171                          & \textbf{-0.111}  & 2.098          & 3.236          & 1.878
                     & 0.378                         & 0.219                          & -0.128           & 1.727          & 2.946          & 1.592          \\
            \textbf{0.8}
                     & \textbf{0.456}                & 0.179                          & -0.129           & \textbf{2.549} & \textbf{3.543} & \textbf{2.197}
                     & \textbf{0.718}                & 0.220                          & -0.118           & \textbf{3.257} & 6.091          & \textbf{2.609} \\
            \textbf{0.9}
                     & 0.415                         & 0.183                          & -0.141           & 2.272          & 2.934          & 2.002
                     & 0.687                         & 0.223                          & \textbf{-0.091}  & 3.087          & \textbf{7.516} & 2.504          \\
            \bottomrule
        \end{tabular}}
\end{table*}

\begin{table*}[t]
    \centering
    \caption{The parameter sensitivity results of parameter label\_t in dataset CSI 300, CSI 500, S\&P 500, and NASDAQ 100.}
    \label{table5}
    \resizebox{\textwidth}{!}{
        \begin{tabular}{c|cccccc|cccccc}
            \toprule
            Metrics  & ARR $\uparrow$                & AVol $\downarrow$              & MDD $\downarrow$ & ASR $\uparrow$ & CR $\uparrow$  & IR $\uparrow$
                     & ARR $\uparrow$                & AVol $\downarrow$              & MDD $\downarrow$ & ASR $\uparrow$ & CR $\uparrow$  & IR $\uparrow$  \\
            \midrule
            Datasets & \multicolumn{6}{c|}{CSI 300}  & \multicolumn{6}{c}{CSI 500}                                                                          \\
            \midrule
            \textbf{2}
                     & 0.107                         & \textbf{0.174}                 & -0.131           & 0.611          & 0.812          & 0.780
                     & 0.179                         & 0.212                          & -0.221           & 0.845          & 0.809          & 0.745          \\
            \textbf{5}
                     & \textbf{0.352}                & 0.226                          & \textbf{-0.127}  & \textbf{1.559} & \textbf{2.776} & \textbf{1.526}
                     & \textbf{0.330}                & \textbf{0.203}                 & -0.198           & \textbf{1.626} & \textbf{1.663} & \textbf{1.382} \\
            \textbf{8}
                     & 0.319                         & 0.220                          & -0.153           & 1.451          & 2.083          & 1.455
                     & 0.293                         & 0.228                          & \textbf{-0.197}  & 1.286          & 1.487          & 1.228          \\
            \midrule
            Datasets & \multicolumn{6}{c|}{S\&P 500} & \multicolumn{6}{c}{NASDAQ 100}                                                                       \\
            \midrule
            \textbf{2}
                     & 0.315                         & \textbf{0.167}                 & \textbf{-0.110}  & 1.884          & 2.854          & 1.720
                     & 0.449                         & 0.230                          & -0.126           & 1.951          & 3.568          & 1.778          \\
            \textbf{5}
                     & \textbf{0.456}                & 0.179                          & -0.129           & \textbf{2.549} & \textbf{3.543} & \textbf{2.197}
                     & \textbf{0.718}                & \textbf{0.220}                 & -0.118           & \textbf{3.257} & \textbf{6.091} & \textbf{2.609} \\
            \textbf{8}
                     & 0.400                         & 0.177                          & -0.126           & 2.255          & 3.175          & 1.992
                     & 0.585                         & 0.229                          & \textbf{-0.117}  & 2.549          & 5.002          & 2.154          \\
            \bottomrule
        \end{tabular}}
\end{table*}

The model integrates several complex components, including the improved GRU, GAT, and multi-head cross-attention mechanisms, which may lead to high computational costs. Specifically, the modified GRU replaces the reset gate in the traditional GRU, and its computational complexity is $O(d \cdot n)$, where $d$ is the hidden state dimension and $n$ is the number of time steps. The computational complexity of the Graph Attention Network (GAT) is $O(|E| \cdot d)$, where $|E|$ is the number of edges in the graph and $d$ is the feature dimension of each node. Since GAT requires calculating attention weights between nodes, its computational complexity is closely related to the sparsity of the graph and the number of nodes. Additionally, the multi-head cross-attention mechanism introduces a computational complexity of $O(n \cdot d_i^2)$ for each attention head, with a total complexity of $O(h \cdot n \cdot d_i^2)$, where $h$ is the number of attention heads and $d_i$ is the dimension of each head. When the number of attention heads $h$ or the dimension $d_i$ is large, the computational cost can increase significantly. To address this challenge and improve the model’s computational efficiency, several optimization strategies can be employed. For instance, parameter sharing and low-rank approximation techniques can be used in both the GRU and multi-head cross-attention mechanisms to reduce the computational burden. Additionally, GAT’s graph structure can be optimized by exploiting sparsity, reducing unnecessary edges, or applying neighborhood sampling to lower the computational complexity. The multi-head attention mechanism can be accelerated by parallelizing the computation of attention heads, reducing overall computation time. Although the model has a high computational cost, its complexity can be effectively controlled through these optimization strategies.

In summary, the experimental results highlight the critical role of integrating relational data and temporal information in stock prediction models. Our proposed model consistently surpasses traditional, deep learning, and reinforcement learning benchmarks across all datasets, achieving superior returns and enhanced risk-adjusted performance metrics. This comprehensive evaluation demonstrates the robustness and efficacy of our approach in capturing the intricate dynamics of financial markets.

\subsection{Parameter Sensitivity}
We conducte an in-depth analysis of the model’s sensitivity to key hyperparameters, including \texttt{judge\_value} (Table \ref{table4}), \texttt{label\_t} (Table \ref{table5}), \texttt{his\_t} (Table \ref{table6}), \texttt{hidden\_size} (Table \ref{table7}), \texttt{gat\_heads} (Table \ref{table8}), and \texttt{num\_hidden\_states} (Table \ref{table9}) across four benchmark datasets. This analysis is conducted to assess the robustness and stability of the model's performance under varying parameter settings. The results demonstrate that the model maintains a high degree of consistency and effectiveness across a broad range of parameter values.

\paragraph{Judge Value Sensitivity:}
The \texttt{judge\_value} parameter, which governs the threshold for filtering graph relationships, exhibits a stable performance profile across all datasets. Specifically, a \texttt{judge\_value} of 0.8 consistently yields the best results, with the model achieving an ARR of 0.352 and an IR of 1.526 on the CSI 300 dataset (Table \ref{table4}). Performance gradually declines as the \texttt{judge\_value} increases beyond this threshold, suggesting that while the model is moderately sensitive to this parameter, it retains robust performance within an optimal range.

\begin{table*}[t]
    \centering
    \caption{The parameter sensitivity results of parameter his\_t in dataset CSI 300, CSI 500, S\&P 500, and NASDAQ 100.}
    \label{table6}
    \resizebox{\textwidth}{!}{
        \begin{tabular}{c|cccccc|cccccc}
            \toprule
            Metrics  & ARR $\uparrow$                & AVol $\downarrow$              & MDD $\downarrow$ & ASR $\uparrow$ & CR $\uparrow$  & IR $\uparrow$
                     & ARR $\uparrow$                & AVol $\downarrow$              & MDD $\downarrow$ & ASR $\uparrow$ & CR $\uparrow$  & IR $\uparrow$  \\
            \midrule
            Datasets & \multicolumn{6}{c|}{CSI 300}  & \multicolumn{6}{c}{CSI 500}                                                                          \\
            \midrule
            \textbf{6}
                     & 0.162                         & 0.216                          & -0.150           & 0.748          & 1.080          & 0.893
                     & 0.228                         & 0.194                          & -0.186           & 1.175          & 1.224          & 1.087          \\
            \textbf{8}
                     & 0.272                         & 0.210                          & -0.140           & 1.291          & 1.943          & 1.329
                     & 0.302                         & 0.224                          & -0.167           & 1.353          & 1.808          & 1.229          \\
            \textbf{10}
                     & \textbf{0.352}                & 0.226                          & -0.127           & \textbf{1.559} & \textbf{2.776} & \textbf{1.526}
                     & \textbf{0.330}                & 0.203                          & -0.198           & \textbf{1.626} & 1.663          & \textbf{1.382} \\
            \textbf{12}
                     & 0.280                         & 0.215                          & -0.169           & 1.299          & 1.655          & 1.347
                     & 0.317                         & 0.214                          & \textbf{-0.109}  & 1.477          & \textbf{2.910} & 1.376          \\
            \textbf{14}
                     & 0.251                         & \textbf{0.193}                 & \textbf{-0.108}  & 1.300          & 2.323          & 1.216
                     & 0.213                         & \textbf{0.180}                 & -0.135           & 1.184          & 1.580          & 1.106          \\
            \midrule
            Datasets & \multicolumn{6}{c|}{S\&P 500} & \multicolumn{6}{c}{NASDAQ 100}                                                                       \\
            \midrule
            \textbf{6}
                     & 0.273                         & \textbf{0.172}                 & -0.125           & 1.588          & 2.186          & 1.494
                     & 0.524                         & 0.229                          & \textbf{-0.099}  & 2.283          & 5.283          & 1.986          \\
            \textbf{8}
                     & 0.363                         & 0.174                          & \textbf{-0.106}  & 2.085          & 3.424          & 1.866
                     & 0.547                         & 0.227                          & -0.106           & 2.406          & 5.162          & 2.066          \\
            \textbf{10}
                     & \textbf{0.456}                & 0.179                          & -0.129           & \textbf{2.549} & \textbf{3.543} & \textbf{2.197}
                     & \textbf{0.718}                & \textbf{0.220}                 & -0.118           & \textbf{3.257} & \textbf{6.091} & \textbf{2.609} \\
            \textbf{12}
                     & 0.406                         & 0.181                          & -0.125           & 2.249          & 3.238          & 1.995
                     & 0.602                         & 0.238                          & -0.115           & 2.529          & 5.248          & 2.131          \\
            \textbf{14}
                     & 0.367                         & 0.175                          & -0.123           & 2.099          & 2.974          & 1.875
                     & 0.603                         & 0.242                          & -0.103           & 2.496          & 5.845          & 2.100          \\
            \bottomrule
        \end{tabular}}
\end{table*}

\begin{table*}[t]
    \centering
    \caption{The parameter sensitivity results of parameter hidden\_size in dataset CSI 300, CSI 500, S\&P 500, and NASDAQ 100.}
    \label{table7}
    \resizebox{\textwidth}{!}{
        \begin{tabular}{c|cccccc|cccccc}
            \toprule
            Metrics  & ARR $\uparrow$                & AVol $\downarrow$              & MDD $\downarrow$ & ASR $\uparrow$ & CR $\uparrow$  & IR $\uparrow$
                     & ARR $\uparrow$                & AVol $\downarrow$              & MDD $\downarrow$ & ASR $\uparrow$ & CR $\uparrow$  & IR $\uparrow$  \\
            \midrule
            Datasets & \multicolumn{6}{c|}{CSI 300}  & \multicolumn{6}{c}{CSI 500}                                                                          \\
            \midrule
            \textbf{16}
                     & 0.179                         & 0.194                          & -0.126           & 0.924          & 1.423          & 1.023
                     & 0.271                         & 0.203                          & -0.203           & 1.337          & 1.334          & 1.244          \\
            \textbf{32}
                     & \textbf{0.352}                & 0.226                          & -0.127           & \textbf{1.559} & \textbf{2.776} & \textbf{1.526}
                     & \textbf{0.330}                & 0.203                          & -0.198           & \textbf{1.626} & 1.663          & \textbf{1.382} \\
            \textbf{64}
                     & 0.283                         & 0.215                          & -0.193           & 1.320          & 1.467          & 1.233
                     & 0.300                         & 0.213                          & -0.120           & 1.408          & \textbf{2.510} & 1.294          \\
            \textbf{128}
                     & 0.204                         & 0.204                          & -0.210           & 0.999          & 0.970          & 1.102
                     & 0.179                         & \textbf{0.197}                 & \textbf{-0.098}  & 0.907          & 1.822          & 0.915          \\
            \textbf{256}
                     & 0.173                         & \textbf{0.180}                 & \textbf{-0.102}  & 0.963          & 1.692          & 1.001
                     & 0.192                         & 0.218                          & -0.216           & 0.881          & 0.889          & 0.898          \\
            \midrule
            Datasets & \multicolumn{6}{c|}{S\&P 500} & \multicolumn{6}{c}{NASDAQ 100}                                                                       \\
            \midrule
            \textbf{16}
                     & 0.411                         & 0.175                          & -0.127           & 2.352          & 3.246          & 2.059
                     & 0.491                         & 0.223                          & -0.106           & 2.203          & 4.616          & 1.943          \\
            \textbf{32}
                     & \textbf{0.456}                & 0.179                          & -0.129           & \textbf{2.549} & \textbf{3.543} & \textbf{2.197}
                     & \textbf{0.718}                & \textbf{0.220}                 & -0.118           & \textbf{3.257} & \textbf{6.091} & \textbf{2.609} \\
            \textbf{64}
                     & 0.406                         & 0.179                          & -0.134           & 2.263          & 3.023          & 1.995
                     & 0.603                         & 0.242                          & \textbf{-0.103}  & 2.496          & 5.845          & 2.100          \\
            \textbf{128}
                     & 0.396                         & 0.177                          & \textbf{-0.104}  & 2.231          & 3.793          & 1.974
                     & 0.511                         & 0.226                          & -0.128           & 2.263          & 4.002          & 1.996          \\
            \textbf{256}
                     & 0.306                         & \textbf{0.172}                 & -0.149           & 1.781          & 2.058          & 1.644
                     & 0.655                         & 0.252                          & -0.143           & 2.604          & 4.584          & 2.179          \\
            \bottomrule
        \end{tabular}}
\end{table*}

\paragraph{Label Time Sensitivity:}
The \texttt{label\_t} parameter, which defines the forecast horizon, also demonstrates strong stability across various datasets. A prediction horizon of 5 days consistently produces the best results, particularly on the NASDAQ 100 dataset, where the model achieves an ARR of 0.718 and an IR of 2.609 (Table \ref{table5}). This indicates that the model effectively captures short to medium-term market trends. The marginal variation in performance across different forecast horizons further highlights the model’s adaptability and resilience in maintaining predictive accuracy.

\paragraph{History Length Sensitivity:}
The \texttt{his\_t} parameter, as shown in Table \ref{table6},  representing the number of historical days considered for prediction, reveals that the model is particularly stable when using a history length of 10 days. This configuration consistently results in the highest ARR and IR across all datasets. The model’s ability to effectively utilize historical data without overfitting or underfitting, even as the history length varies, underscores its robustness in learning from past patterns.

\paragraph{Hidden Size Sensitivity:}
The hidden size, which defines the dimensionality of the model's internal representations, plays a significant role in shaping the model’s performance. Our analysis identifies a hidden size of 32 (Table \ref{table7}) as the optimal configuration. The model's performance remains stable across different hidden size values, demonstrating its capability to balance model complexity with predictive accuracy, and effectively handle the diverse characteristics of financial data.

\paragraph{GAT Heads Sensitivity:}
The number of graph attention heads (\texttt{gat\_heads}) also significantly impacts the model’s ability to capture the complex dependencies in the data. Our experiments show that using 4 attention heads consistently results in the highest performance, with the model achieving an ARR of 0.352 and an IR of 1.526 on the CSI 300 dataset (Table \ref{table8}). This stability across different settings of \texttt{gat\_heads} reflects the model’s robustness in learning the intricate relationships between financial assets.

\paragraph{Num Hidden States Sensitivity:}
The number of hidden states (\texttt{num\_hidden\_states}) represents the market's latent dynamics. From our experiments (Table \ref{table9}), we find that using 4 or 8 hidden states yields the best performance across all datasets. Specifically, for the CSI 300 dataset, setting \texttt{num\_hidden\_states} to 8 results in the highest ARR of 0.356 and an IR of 1.533. Similarly, for the S\&P 500 and NASDAQ 100 datasets, 4 hidden states lead to superior performance, with the model achieving an ARR of 0.456 and an IR of 2.197 on the S\&P 500, and an ARR of 0.718 and an IR of 2.609 on the NASDAQ 100. This suggests that a moderate number of hidden states is optimal for capturing the underlying market structures, while too few or too many hidden states can result in a decline in performance, as seen with the configurations of 2 and 16 hidden states.

\begin{table*}[t]
    \centering
    \caption{The parameter sensitivity results of parameter gat\_heads in dataset CSI 300, CSI 500, S\&P 500, and NASDAQ 100.}
    \label{table8}
    \resizebox{\textwidth}{!}{
        \begin{tabular}{c|cccccc|cccccc}
            \toprule
            Metrics  & ARR $\uparrow$                & AVol $\downarrow$              & MDD $\downarrow$ & ASR $\uparrow$ & CR $\uparrow$  & IR $\uparrow$
                     & ARR $\uparrow$                & AVol $\downarrow$              & MDD $\downarrow$ & ASR $\uparrow$ & CR $\uparrow$  & IR $\uparrow$  \\
            \midrule
            Datasets & \multicolumn{6}{c|}{CSI 300}  & \multicolumn{6}{c}{CSI 500}                                                                          \\
            \midrule
            \textbf{1}
                     & 0.253                         & \textbf{0.203}                 & \textbf{-0.110}  & 1.250          & 2.300          & 1.215
                     & 0.200                         & \textbf{0.195}                 & -0.201           & 1.023          & 0.992          & 0.961          \\
            \textbf{2}
                     & 0.285                         & 0.210                          & -0.150           & 1.359          & 1.907          & 1.383
                     & 0.296                         & 0.229                          & \textbf{-0.130}  & 1.290          & \textbf{2.270} & 1.199          \\
            \textbf{4}
                     & \textbf{0.352}                & 0.226                          & -0.127           & \textbf{1.559} & \textbf{2.776} & \textbf{1.526}
                     & \textbf{0.330}                & 0.203                          & -0.198           & \textbf{1.626} & 1.663          & \textbf{1.382} \\
            \textbf{8}
                     & 0.239                         & 0.211                          & -0.133           & 1.133          & 1.794          & 1.084
                     & 0.300                         & 0.230                          & -0.198           & 1.305          & 1.519          & 1.218          \\
            \midrule
            Datasets & \multicolumn{6}{c|}{S\&P 500} & \multicolumn{6}{c}{NASDAQ 100}                                                                       \\
            \midrule
            \textbf{1}
                     & 0.322                         & 0.181                          & -0.133           & 1.781          & 2.414          & 1.640
                     & 0.686                         & 0.235                          & -0.115           & 2.918          & 5.952          & 2.372          \\
            \textbf{2}
                     & 0.385                         & 0.183                          & -0.147           & 2.102          & 2.623          & 1.876
                     & 0.666                         & 0.239                          & \textbf{-0.111}  & 2.794          & 6.028          & 2.287          \\
            \textbf{4}
                     & \textbf{0.456}                & 0.179                          & -0.129           & \textbf{2.549} & 3.543          & 2.197
                     & \textbf{0.718}                & \textbf{0.220}                 & -0.118           & \textbf{3.257} & \textbf{6.091} & \textbf{2.609} \\
            \textbf{8}
                     & 0.440                         & \textbf{0.162}                 & \textbf{-0.116}  & 2.718          & \textbf{3.807} & \textbf{2.345}
                     & 0.588                         & 0.226                          & -0.127           & 2.604          & 4.619          & 2.194          \\
            \bottomrule
        \end{tabular}}
\end{table*}

\begin{table*}[t]
    \centering
    \caption{The parameter sensitivity results of parameter num\_hidden\_states in dataset CSI 300, CSI 500, S\&P 500, and NASDAQ 100.}
    \label{table9}
    \resizebox{\textwidth}{!}{
        \begin{tabular}{c|cccccc|cccccc}
            \toprule
            Metrics  & ARR $\uparrow$                & AVol $\downarrow$              & MDD $\downarrow$ & ASR $\uparrow$ & CR $\uparrow$  & IR $\uparrow$
                     & ARR $\uparrow$                & AVol $\downarrow$              & MDD $\downarrow$ & ASR $\uparrow$ & CR $\uparrow$  & IR $\uparrow$  \\
            \midrule
            Datasets & \multicolumn{6}{c|}{CSI 300}  & \multicolumn{6}{c}{CSI 500}                                                                          \\
            \midrule
            \textbf{2}
                     & 0.274                         & 0.207                          & \textbf{-0.114}  & 1.320          & 2.398          & 1.361
                     & 0.294                         & 0.242                          & -0.256           & 1.215          & 1.147          & 1.133          \\
            \textbf{4}
                     & 0.352                         & 0.226                          & -0.127           & 1.559          & 2.776          & 1.526
                     & 0.330                         & 0.203                          & -0.198           & 1.626          & 1.663          & 1.382          \\
            \textbf{8}
                     & \textbf{0.356}                & 0.222                          & -0.125           & \textbf{1.603} & \textbf{2.846} & \textbf{1.533}
                     & \textbf{0.381}                & 0.202                          & \textbf{-0.108}  & \textbf{1.888} & \textbf{3.531} & \textbf{1.658} \\
            \textbf{16}
                     & 0.199                         & \textbf{0.199}                 & -0.153           & 0.998          & 1.301          & 1.060
                     & 0.250                         & \textbf{0.198}                 & -0.134           & 1.262          & 1.865          & 1.159          \\
            \midrule
            Datasets & \multicolumn{6}{c|}{S\&P 500} & \multicolumn{6}{c}{NASDAQ 100}                                                                       \\
            \midrule
            \textbf{2}
                     & 0.420                         & \textbf{0.176}                 & -0.135           & 2.382          & 3.120          & 2.084
                     & 0.618                         & 0.235                          & \textbf{-0.107}  & 2.629          & 5.755          & 2.196          \\
            \textbf{4}
                     & \textbf{0.456}                & 0.179                          & -0.129           & \textbf{2.549} & 3.543          & \textbf{2.197}
                     & \textbf{0.718}                & \textbf{0.220}                 & -0.118           & \textbf{3.257} & \textbf{6.091} & \textbf{2.609} \\
            \textbf{8}
                     & 0.435                         & 0.232                          & \textbf{-0.114}  & 1.875          & \textbf{3.803} & 1.686
                     & 0.506                         & 0.229                          & -0.112           & 2.210          & 4.511          & 1.933          \\
            \textbf{16}
                     & 0.341                         & 0.182                          & -0.139           & 1.876          & 2.461          & 1.704
                     & 0.523                         & 0.226                          & -0.108           & 2.320          & 4.856          & 2.009          \\
            \bottomrule
        \end{tabular}}
\end{table*}

\paragraph{Overall Observations:}
Across all evaluated parameters, the model consistently exhibits a high level of stability and resilience, with minimal performance fluctuations under varying settings. This robustness suggests that the model is well-regularized and capable of maintaining strong predictive accuracy across a wide range of hyperparameter configurations. Such stability is critical in practical applications, where models are required to operate under varying market conditions and data distributions. The consistency of results across different datasets and parameter configurations further underscores the reliability and utility of the proposed model in financial forecasting tasks.

\subsection{Ablation Study}
In this section, we conduct a comprehensive ablation study to evaluate the individual contributions of various components of our model. The model is systematically divided into four distinct modules: Use Improved GRU to Capture Temporal Features~(I), Use GAT to Capture Cross-sectional Features~(II), Use Multi-head Cross-attention to Capture Latent State Features~(III), Model Prediction and Loss Calculation Layer~(IV). Detailed experimental results are presented in Tables \ref{table10} and \ref{table11}.

\begin{table*}[t]
    \centering
    \caption{The ablation study results in dataset CSI 300 and CSI 500.}
    \label{table10}
    \resizebox{\textwidth}{!}{
        \begin{tabular}{c|cccccc|cccccc}
            \toprule
            Datasets & \multicolumn{6}{c|}{CSI 300} & \multicolumn{6}{c}{CSI 500}                                                                       \\
            Model    & ARR $\uparrow$               & AVol $\downarrow$           & MDD $\downarrow$ & ASR $\uparrow$ & CR $\uparrow$  & IR $\uparrow$
                     & ARR $\uparrow$               & AVol $\downarrow$           & MDD $\downarrow$ & ASR $\uparrow$ & CR $\uparrow$  & IR $\uparrow$  \\
            \midrule
            \textbf{I + II}
                     & 0.076                        & 0.172                       & -0.214           & 0.441          & 0.354          & 0.488
                     & 0.151                        & 0.178                       & -0.169           & 0.850          & 0.891          & 0.984          \\
            \textbf{I + II + III}
                     & 0.155                        & 0.166                       & -0.170           & 0.934          & 0.911          & 0.957
                     & 0.195                        & 0.219                       & -0.156           & 0.890          & 1.247          & 0.882          \\
            \textbf{I + II + IV}
                     & 0.210                        & 0.201                       & -0.183           & 1.039          & 1.145          & 1.113
                     & 0.222                        & 0.191                       & -0.166           & 1.166          & 1.335          & 1.191          \\
            \textbf{I + III}
                     & -0.069                       & \textbf{0.160}              & -0.215           & -0.435         & -0.322         & -0.428
                     & 0.090                        & \textbf{0.160}              & -0.168           & 0.560          & 0.534          & 0.548          \\
            \textbf{I + III + IV}
                     & 0.110                        & 0.239                       & -0.199           & 0.459          & 0.554          & 0.551
                     & 0.192                        & 0.243                       & -0.197           & 0.793          & 0.976          & 0.807          \\
            \textbf{II + III}
                     & 0.151                        & 0.178                       & -0.169           & 0.850          & 0.891          & 0.984
                     & 0.244                        & 0.216                       & \textbf{-0.154}  & 1.129          & 1.580          & 1.046          \\
            \textbf{II + III + IV}
                     & 0.287                        & 0.211                       & \textbf{-0.115}  & 1.360          & 2.498          & 1.390
                     & 0.244                        & 0.200                       & -0.187           & 1.223          & 1.304          & 1.150          \\
            \textbf{MCI-GRU}
                     & \textbf{0.352}               & 0.226                       & -0.127           & \textbf{1.559} & \textbf{2.776} & \textbf{1.526}
                     & \textbf{0.330}               & 0.203                       & -0.198           & \textbf{1.626} & \textbf{1.663} & \textbf{1.382} \\
            \bottomrule
        \end{tabular}}
\end{table*}

\begin{table*}[t]
    \centering
    \caption{The ablation study results in dataset NASDAQ 100 and S\&P 500.}
    \label{table11}
    \resizebox{\textwidth}{!}{
        \begin{tabular}{c|cccccc|cccccc}
            \toprule
            Datasets & \multicolumn{6}{c|}{S\&P 500} & \multicolumn{6}{c}{NASDAQ 100}                                                                       \\
            Model    & ARR $\uparrow$                & AVol $\downarrow$              & MDD $\downarrow$ & ASR $\uparrow$ & CR $\uparrow$  & IR $\uparrow$
                     & ARR $\uparrow$                & AVol $\downarrow$              & MDD $\downarrow$ & ASR $\uparrow$ & CR $\uparrow$  & IR $\uparrow$  \\
            \midrule
            \textbf{I + II}
                     & 0.305                         & 0.179                          & \textbf{-0.120}  & 1.703          & 2.552          & 1.683
                     & 0.550                         & \textbf{0.194}                 & -0.101           & 2.841          & 5.461          & 2.483          \\
            \textbf{I + II + III}
                     & 0.338                         & 0.194                          & -0.178           & 1.736          & 1.897          & 1.691
                     & 0.648                         & 0.238                          & -0.112           & 2.721          & 5.807          & 2.248          \\
            \textbf{I + II + IV}
                     & 0.221                         & 0.194                          & -0.155           & 1.143          & 1.428          & 1.240
                     & 0.444                         & 0.195                          & -0.116           & 2.278          & 3.822          & 2.109          \\
            \textbf{I + III}
                     & 0.332                         & 0.163                          & -0.121           & 2.040          & 2.736          & 1.803
                     & 0.342                         & 0.194                          & -0.152           & 1.766          & 2.245          & 1.684          \\
            \textbf{I + III + IV}
                     & 0.212                         & \textbf{0.156}                 & -0.135           & 1.361          & 1.578          & 1.322
                     & 0.470                         & 0.243                          & -0.152           & 1.933          & 3.095          & 1.728          \\
            \textbf{II + III}
                     & 0.390                         & 0.176                          & -0.151           & 2.214          & 2.579          & 1.968
                     & 0.524                         & 0.226                          & -0.109           & 2.320          & 4.771          & 2.026          \\
            \textbf{II + III + IV}
                     & 0.416                         & 0.181                          & -0.157           & 2.296          & 2.647          & 2.018
                     & 0.564                         & 0.235                          & \textbf{-0.100}  & 2.405          & 5.646          & 2.035          \\
            \textbf{MCI-GRU}
                     & \textbf{0.456}                & 0.179                          & -0.129           & \textbf{2.549} & \textbf{3.543} & \textbf{2.197}
                     & \textbf{0.718}                & 0.220                          & -0.118           & \textbf{3.257} & \textbf{6.091} & \textbf{2.609} \\
            \bottomrule
        \end{tabular}}
\end{table*}

The results presented in Tables \ref{table10} and \ref{table11} reveal several critical insights. First, the integration of the I and II resulted in a moderate improvement in performance metrics across all datasets, indicating that these components play a significant role in capturing both temporal and relational dependencies within the data. However, the addition of the III further amplified the model's performance, suggesting that the incorporation of market-wide latent states provides a more comprehensive understanding of broader market dynamics. This enhancement underscores the importance of integrating diverse levels of information to effectively model complex financial relationships.

In comparing models that incorporate the loss calculation layer, the sub-model configurations~(I+II+IV) and (I+III+IV) exhibited varying degrees of improvement, with a particularly notable enhancement in the CSI 300 and CSI 500 datasets. These results suggest that the inclusion of the loss calculation layer significantly refines the model's predictions by further processing the concatenated feature vectors. Specifically, the configuration~(II+III+IV) demonstrated substantial performance gains, particularly in ARR and ASR metrics, underscoring the effectiveness of combining the GAT layer, market hidden states, and an optimized loss calculation mechanism to boost predictive accuracy.

In conclusion, the ablation study demonstrates that the proposed model components are not only complementary but their integration substantially enhances predictive performance. This analysis underscores the critical importance of leveraging both temporal and relational information, in conjunction with latent market states, to achieve precise and reliable financial forecasting.

\subsection{Case studies}
This section offers a comprehensive explanation of the practical deployment of our model within EMoney Inc.'s real-world algorithmic trading platform, demonstrating its robust adaptability to dynamic financial environments. The model is trained on a monthly basis, generating daily predictions immediately after the close of each trading session. These predictions serve as the foundation for executing trading strategies in the initial half-hour of the next trading day. This early execution period leverages the model's predictions to capitalize on short-term market movements. The strategies we implement are tailored specifically to the CSI 300, CSI 500 and CSI 1000 stock pools, with optimization processes designed to blend the distinct characteristics of both indices, enhancing the model’s overall effectiveness across diverse market conditions.

\begin{figure}[htb]
    \centering
    \subfigure[CSI 300 strategy backtest performance.]{\label{figure2_1}
        \includegraphics[width=0.9\textwidth,height=0.23\textwidth]{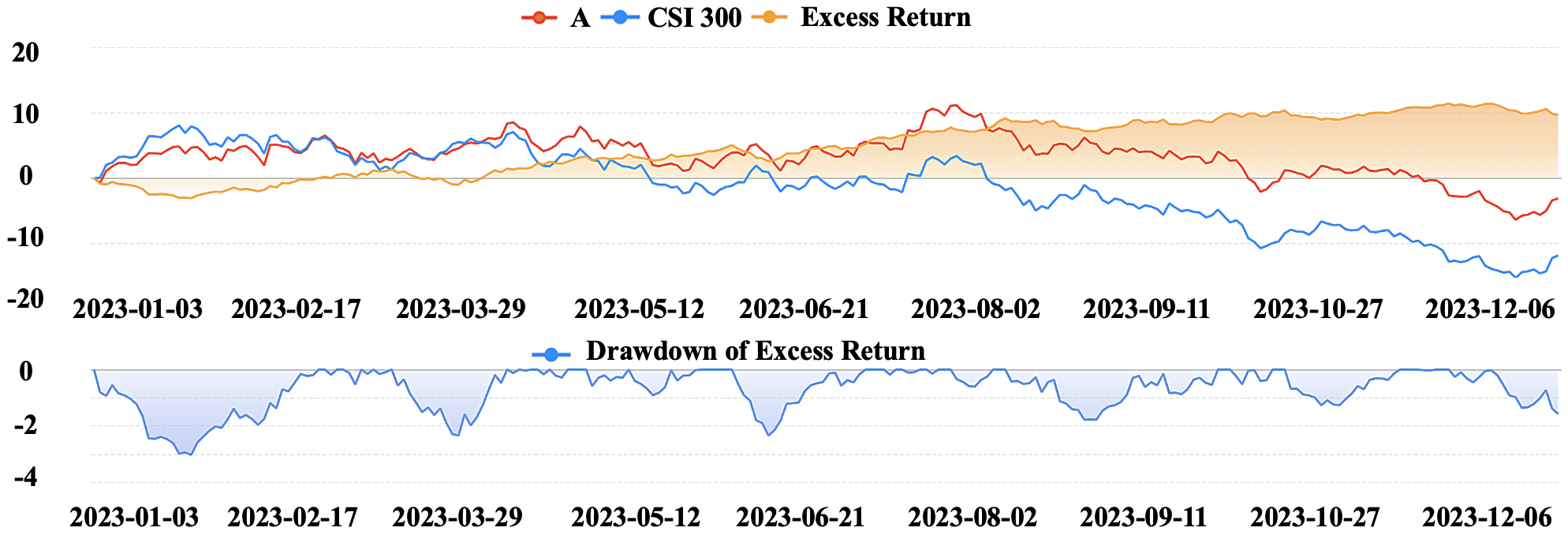}}
    \subfigure[CSI 500 strategy backtest performance.]{\label{figure2_2}
        \includegraphics[width=0.9\textwidth,height=0.23\textwidth]{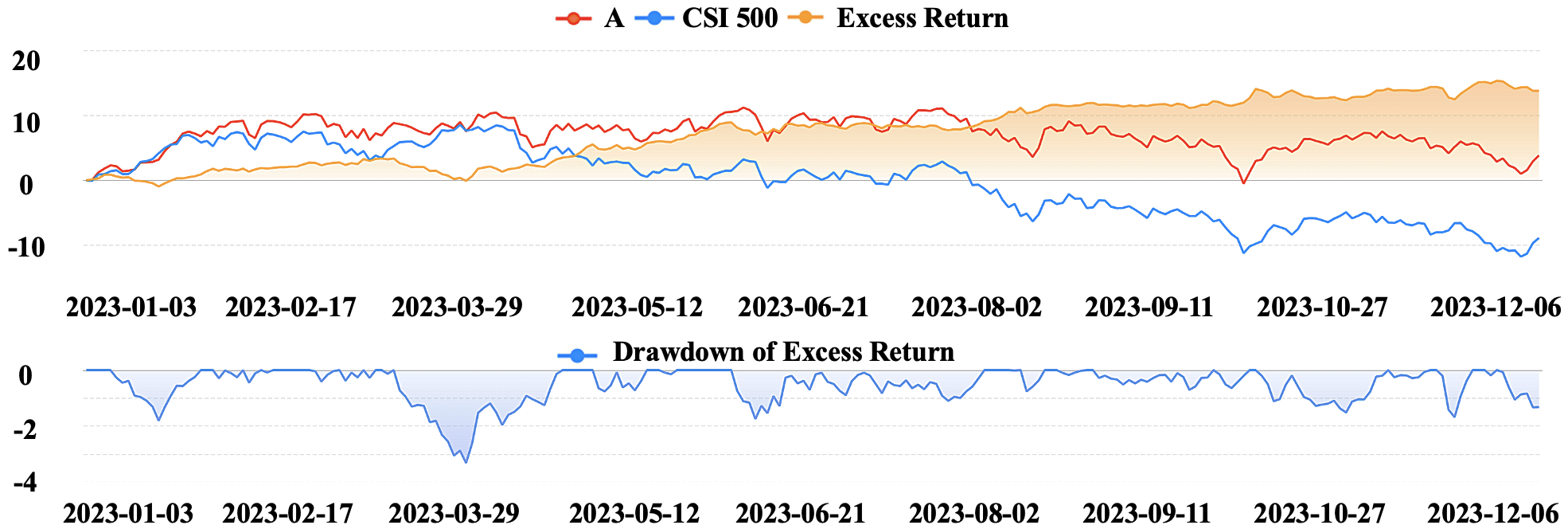}}
    \subfigure[CSI 1000 strategy backtest performance.]{\label{figure2_3}
        \includegraphics[width=0.9\textwidth,height=0.23\textwidth]{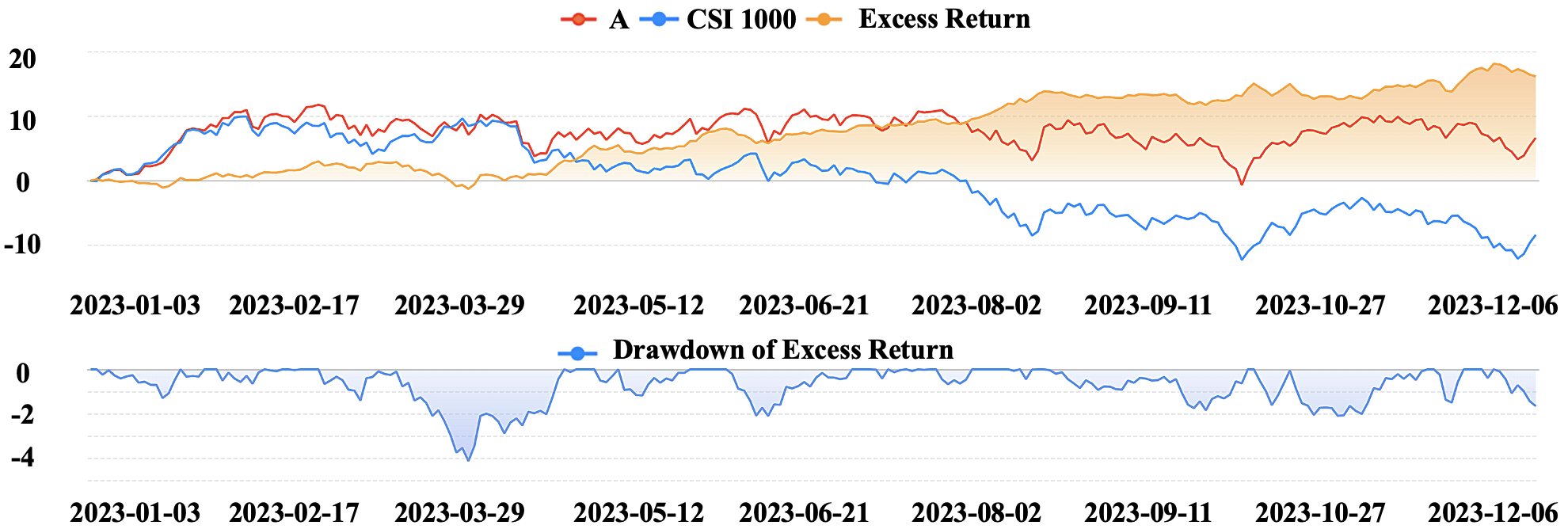}}
    \caption{The performance of the strategy backtest.}
    \label{figure2}
\end{figure}

Figure \ref{figure2} visualizes the performance of these strategies across different scenarios. In Figure \ref{figure2_1},  \ref{figure2_2} and \ref{figure2_3}, the red curve illustrates the model’s absolute returns, which reflect the actual profitability of the trades based on our model's predictions. In comparison, the blue curve shows the performance of the CSI 300, CSI 500, and CSI 1000 indices, representing the overall market return trends. The yellow curve denotes the excess returns, highlighting the model’s ability to generate returns beyond the market average. Throughout a one-year period, the results consistently demonstrate that the model’s strategies significantly outperform the market indices, showcasing its superior predictive power and strategic effectiveness. Moreover, the lower part of Figure \ref{figure2_1},  \ref{figure2_2} and \ref{figure2_3} provide an analysis of the excess return drawdown rate, a critical metric for assessing the model’s risk management capabilities. The drawdown rate measures the extent to which excess returns decline from their peak to their lowest point, reflecting the model’s ability to mitigate risk during market downturns. The model exhibits exceptional risk control, maintaining a consistently low drawdown rate, with the worst-case scenario showing a reduction of just about 5\%. This low drawdown rate indicates that the model not only prioritizes return generation but also incorporates robust risk mitigation mechanisms, ensuring a balance between profit-seeking and risk aversion. This capability is particularly important in real-world trading environments where minimizing losses during periods of volatility is crucial for long-term success.

\subsection{Limitations of the model}
Although the proposed MCI-GRU model has demonstrated excellent performance in multiple experiments and has been successfully applied in a fund management company, we have not discussed its potential limitations in detail. In terms of scalability, as the dataset size increases, the training time and computational resource consumption of the MCI-GRU model may grow, especially when dealing with a large number of stocks or multiple markets. While we have achieved good results on the Chinese and U.S. stock market datasets, the training and inference efficiency of the model may be impacted as the data scale expands. Future work will focus on improving scalability through methods such as optimizing computational graphs, model compression, or distributed training.

Regarding robustness to market volatility, although the model performs well on historical data, its robustness under extreme market fluctuations still needs further validation. In real-world applications, severe market volatility may lead to model prediction failure. Therefore, we plan to explore the model's adaptability in different market environments and consider incorporating risk management mechanisms to enhance the model's robustness to market volatility.

As for sensitivity to hyperparameter tuning, although we have selected appropriate hyperparameters in our experiments and validated the model’s robustness to some extent, MCI-GRU may still be highly sensitive to certain hyperparameters (such as learning rate, number of attention heads, etc.). To reduce this sensitivity, future research will explore adaptive hyperparameter tuning methods or use techniques like Bayesian optimization to further improve the model’s generalization ability.

\section{Conclusion}
In this paper, we present a novel stock prediction model, MCI-GRU, which integrates a multi-head cross-attention mechanism and the improved GRU architecture to address the challenges of capturing complex temporal and relational dependencies in stock data. By replacing the reset gate in the traditional GRU with an attention mechanism, the model significantly improves its capacity to selectively utilize historical time series data. Additionally, the incorporation of the GAT enables the extraction of cross-sectional features, while the multi-head cross-attention mechanism captures latent market states that influence stock behavior. Extensive experiments conducted on both Chinese and U.S. stock market datasets demonstrate that MCI-GRU outperforms existing state-of-the-art methods across various performance metrics. Moreover, the model has been successfully implemented in a real-world fund management company, showcasing its practical applicability.

In the current work, we employed static graph methods to simplify the model implementation, focusing on static correlation feature extraction. However, dynamic graphs are more effective at capturing real-time market changes, especially in the dynamic financial market. Future research will introduce dynamic graph methods to improve feature extraction quality and enhance the model's adaptability to market changes. Additionally, although MCI-GRU has performed well in several experiments and has been successfully applied in a fund management company, we recognize challenges in scalability, robustness to market volatility, and sensitivity to hyperparameters. Future work will explore methods such as optimizing computational graphs, distributed training, and adaptive hyperparameter tuning to further improve the model's efficiency and robustness.

\section*{Acknowledgements}
This work was supported in part by the National Key Research and Development Program of China under Grant 2022YFB4501704,
the National Science Foundation of China under Grants 62222312,
and the Shanghai Science and Technology Innovation Action Plan Project under Grant 22511100700 and 22YS1400600.


\bibliographystyle{elsarticle-num}
\bibliography{elsarticle-template-num}

\end{document}